\documentclass [11pt,a4paper] {article}

\usepackage[T1]{fontenc}
\usepackage{lmodern}

\usepackage[cp1252]{inputenc}
\usepackage{amssymb}
\usepackage{amsmath}
\usepackage{amsfonts,amssymb}
\usepackage[dvips]{graphicx}
\usepackage{bbm}
\usepackage{enumerate}
\usepackage{amsthm}
\usepackage{cancel}

\DeclareMathAlphabet{\mathpzc}{OT1}{pzc}{m}{it}

\def\SmallColSep{\setlength{\arraycolsep}{1pt}}

\newcommand*\rfrac[2]{{}^{#1}\!\!/\!_{#2}}

\begin{document}

\title{Wave–particle duality and the objectiveness of “true” and “false”}

\author{Arkady Bolotin\footnote{$Email: arkadyv@bgu.ac.il$\vspace{5pt}} \\ \emph{Ben-Gurion University of the Negev, Beersheba (Israel)}}

\maketitle

\begin{abstract}\noindent The traditional analysis of the basic version of the double-slit experiment leads to the conclusion that wave-particle duality is a fundamental fact of nature. However, such a conclusion means to imply that we are not only required to have two contradictory pictures of reality but also compelled to abandon the objectiveness of the truth values, “true” and “false”. Yet, even if we could accept wave-like behavior of quantum particles as the best explanation for the build-up of an interference pattern in the double-slit experiment, without the objectivity of the truth values we would never have certainty regarding any statement about the world. The present paper discusses ways to reconcile the correct description of the double-slit experiment with the objectiveness of “true” and “false”.\\\\

\end{abstract}

\section{Introduction}  

\noindent Wave–particle duality is the concept of quantum mechanics, which holds that every quantum particle can be described as either a particle or a wave \cite{Greiner}. This dualism is a direct consequence of Niels Bohr’s principle of complementarity which states that having precise knowledge of one of two experimental outcomes complementing each other prevents us from obtaining complete information about the remaining one \cite{Bohr}. Though wave–particle dualism was proposed long ago, an explanation of its meaning has not been agreed yet, and it persists in being perplexing to the mind. As an illustration of an unfading interest to the question of “particle” versus “wave” and its relevancy in our time, see papers \cite{Scully} and \cite{Menzel}, just to name a few.\\

\noindent To make plain the perplexing character of wave–particle duality, let us consider the basic version of the Young's double-slit experiment: Emitted one at a time, quantum particles (like photons or electrons) hit a plate pierced by two slits (labeled 1 and 2), which are located along the $x$-axis at $x=0$ and $x=d$, respectively, and are afterwards observed on a screen behind the plate.\\

\noindent Let us examine the following statement: {\guillemotleft}In the double-slit experiment, the quantum particle passes through one or the other slit, but not both{\guillemotright} (to set statements off from the rest of the text, in this paper they are inserted in the double angle quotation marks). For brevity, let the above statement be denoted by the capital letter $C$.\\

\noindent On the word of Feynman \cite{Feynman}, if one has a piece of apparatus able to determine whether a quantum particle goes through the slit 1 or the slit 2 (called a which-way detector), then one can say that the particle goes through either the slit 1 or the slit 2 and so one can say that the statement $C$ is true. But, when both slits are open and there is no which-way detector, an interference pattern builds up slowly on the screen as more particles go through the slits. In that case, one may not say that each quantum particle passes through either the slit 1 or the slit 2, and, correspondingly, one may not say that the statement $C$ is true. Providing $C$ may be either true or false, this last means that the statement $C$ is false.\\

\noindent In this way, the concept of wave–particle duality brings in the dual valuation for the statement $C$. Namely, in case the quantum particle is described as a classically defined particle, $C$ is true, but if the quantum particle is described as a classically defined wave, $C$ becomes false.\\

\noindent However, the problem is that the character of the statement $C$ is quite different from that of \emph{a contingent statement} which may be true in one instance but false in another. As a matter of fact, if a which-way detector is present in the experiment, the statement $C$ is \emph{a tautology}, i.e., $C$ is true in every instance (here “instance” is understood as a recording of the particle’s position on the screen). Put differently, $C$ is true regardless of the truths and falsities of the contingent statements {\guillemotleft}The quantum particle passes through the slit 1{\guillemotright} denoted $P_1$ and {\guillemotleft}The quantum particle passes through the slit 2{\guillemotright} denoted $P_2$. But as soon as the which-way detector is out (and both slits are open), the statement $C$ (that can be written as {\guillemotleft}$P_1$ or $P_2$, but not, $P_1$ and $P_2${\guillemotright}) makes a transition from a tautology to \emph{a contradiction} since then there are no instances wherein $C$ could fail to be false.\enlargethispage{\baselineskip}\\

\noindent A transformation of a tautology into a contradiction (and vice versa) would be possible if “true” and “false” were to be not objective (absolute) but relative to experience. To be sure, suppose that “true” and “false” were dependent on the act of observing the quantum particle in the double-slit experiment. Then, the type of observation would determine which of the properties of the quantum particle – pertaining to a particle or to a wave – would show up in the experiment and, hence, what kind of a logical statement would be $C$ – a tautology or a contradiction.\\

\noindent More to the point, let both slits be open and suppose that a which-way detector is brought in the experiment sometime. Then, one finds that even if the statement $C$ initially is a contradiction, it may nonetheless turn out to be a tautology at a later time. This would be equivalent to saying that “false” can be converted into “true” when the observation is made or when “false'' is subject to empirical testing. Clearly, this could happen only if the truth values were subjective.\\

\noindent On the other hand, if “true” and “false” were to be nonabsolute, we would never have certainty regarding any statement about the world – including statements grounded in meanings, independent of matters of fact. Take, for example, the mathematical statement {\guillemotleft}2+2=5{\guillemotright}. It is false solely by definition, that is, we need not consult experience to determine whether {\guillemotleft}2+2=5{\guillemotright} is false or not. Moreover, even though this statement can be falsified by experience, it is not grounded in experience. This means that the aforesaid statement is \emph{necessarily false} and so we can be certain that it was false in the past and will remain false at any future moment, regardless of the limits of our present knowledge or our powers of theoretical understanding. However, were “false” to be subjective, the statement {\guillemotleft}2+2=5{\guillemotright} would be \emph{empirically false}, that is, the falsity of {\guillemotleft}2+2=5{\guillemotright} would be contingent on the observation of facts. As a result, we could not rule out an occasion when this statement was verified by experience.\\

\noindent Hence, the wave–particle duality appears to lead to the conclusion that the truths and falsities of logic and mathematics need confirmation by observations.\\

\noindent Since this conclusion is controvertible at best, the puzzle is, then, how to bring together the successful description of the double-slit experiment with the objectiveness of “true” and “false”. The present paper seeks the answer to this puzzle.\\

\section{The motivation}  

\noindent The attentive reader may have noticed that the validity of dual valuation for the statement $C$, pointed out in the previous section, is based on the following two conditions:\smallskip

\begin{description}
\item [(a)] a truth value of the complex statement {\guillemotleft}$P_1$ or $P_2$, but not, $P_1$ and $P_2${\guillemotright} entirely depends on truth values of its constituent statements, i.e., $P_1$ and $P_2$,
\item [(b)] a relation between the set of statements and the set of the truth values (that has just two members, “true” and “false”) is a total function (as a result, every statement is either true or false but not both, and it is not the case that a statement is not true and not false).
\end{description}

\noindent So, to dismiss dual valuation for $C$ (and thus preserve the objectiveness of “true” and “false”), either one (or all) of the above conditions must be revised.\\

\noindent As is well known, it is bivalent truth-functional propositional logic (usually called \emph{classical logic}) which takes for granted that truth values of complex statements are defined by truth values of their constituent statements. Therefore, a revision of the condition \textbf{(a)} entails nothing less than a modification of classical logic.\\

\noindent Potentially, the laws of classical logic can be modified in various ways. One of such modifications (that replaces the classical logical connectives by different ones inspired by the lattice-oriented operations) is \emph{quantum logic} proposed by Birkhoff and von Neumann \cite{Birkhoff} in 1936. Since that time, other forms of quantum logics have been developed, a few of them are: \emph{a dynamic quantum logic} \cite{Baltag}, \emph{exogenous quantum propositional logic} \cite{Mateus}, and \emph{a categorical quantum logic} \cite{Abramsky} (more can be found in \cite{Pavicic}).\\

\noindent Another modification of classical logic is \emph{the calculus of partial propositional functions} introduced in \cite{Kochen} and systematically studied in \cite{Specker}. This modification takes into account that arbitrary statements may be pairwise ``incompatible'' and thus non-connectable. According to the calculus of partial propositional functions, the complex statements {\guillemotleft}$X$ or not-$X${\guillemotright} and {\guillemotleft}$X$ and not-$X${\guillemotright} are always true and false, respectively, even though their constituent statements may lack a truth value. What is more, this lack of a truth value can be considered responsible for the interference pattern in the double-slit experiment \cite{Kochen15}.\\

\noindent On the other hand, no modification can change the fact that propositional logic – as a mathematical model that allows us to reason about the truth or falsity of expressions constructed from simple statements (i.e., ones that are not linked by logical connectives) – cannot see inside those statements. For example, consider two simple statements: {\guillemotleft}In the double-slit experiment, both slits are open{\guillemotright} and {\guillemotleft}In the double-slit experiment, only one slit is open{\guillemotright}. Let the letters $S_{\text{both}}$ and $S_{\text{one}}$ denote the said statements in the order given. Propositional logic sees $S_{\text{both}}$ and $S_{\text{one}}$ as indivisible entities, and from that reason, the parts of $S_{\text{both}}$ and $S_{\text{one}}$ concerning the number of open slits cannot be taken into consideration. This implies that it is impossible to distinguish {\guillemotleft}$S_{\text{both}}$ and $P_n${\guillemotright} from {\guillemotleft}$S_{\text{one}}$ and $P_n${\guillemotright} (where $n$ is 1 or 2) and, hence, to decide which of these complex statements may lack a truth value by remaining exclusively within the frame of propositional logic (or for that matter, any of its modifications including the calculus of partial propositional functions and whichever form of quantum logic).\\

\noindent But rather than exploring more complicated branches of logic to address this problem, one may prefer to keep up the laws of classical logic and use an alternative approach, in which an assignment of truth values to statements about quantum systems directly accesses the mathematical formalism of quantum theory. This approach will be presented in the next sections.\\

\section{Truth values of statements about the double-slit experiment}  

\noindent Let us start by recalling that a simple statement of a mathematical relation (as equality or inequality) between meaningful expressions (i.e., symbols or combinations of symbols representing a value, a function, an object or the like) is called \emph{an atomic mathematical statement} \cite{Church, Mendelson}. A statement like this can be classified as \emph{analytic} since its truth and falsity depend solely on the meaning of its terms. A simple \emph{synthetic} (i.e., not analytic) statement, which affirms or denies something meaningful about the world and is capable of being true or false, is called \emph{an atomic proposition} \cite{Klement}.\\

\noindent More complex mathematical statements (or propositions) are called molecular ones. They are built up out of atomic components via logical connectives, for example, $\sqcup$ (logical disjunction), $\sqcap$ (logical conjunction) and $\neg$ (logical negation).\\ 

\noindent In the paper, an atomic mathematical statement is denoted by a lowercase letter (which may contain sub- and superscript characters), for example, $\text{\guillemotleft}\mkern1mu{3\le4}\mkern1mu\text{\guillemotright}\mkern-2mu=a_1$, while an atomic proposition is denoted by an uppercase letter (which also may contain sub- and superscript characters), e.g., $\text{\guillemotleft}\mkern1mu\text{Today is Sunday}\mkern1mu\text{\guillemotright}\mkern-2mu=A$, where the combination of $\text{\guillemotleft}\mkern-1mu\dots\mkern-1.5mu\text{\guillemotright}$ and $=$ stands for ``\dots is denoted by \dots''.\\

\noindent Correspondingly, the atomic proposition asserting that in the double-slit experiment the quantum particle passes through the particular slit -- 1 or 2 -- can be presented in the following way:\smallskip

\begin{equation}  
   n
   \in
   \{1,2\}
   \mkern-3.3mu
   :
   \mkern4mu
   \text{\guillemotleft}
   \mkern1mu
   \text{The quantum particle passes through the slit}
   \mkern4mu
   n
   \mkern1mu
   \text{\guillemotright}
   \mkern-2mu
   =
   P_n
   \;\;\;\;  .
\end{equation}
\smallskip

\noindent Suppose that at the moment when the quantum particle comes out from the double-slit plate, its state is described by $\Psi\mkern-2mu\left( x\right)$, the complex scalar function of $x$ (whose codomain contains more than the value 0). Providing $\Psi\mkern-2mu\left( x\right)$ is given, one can consider the atomic mathematical statement posed as\smallskip

\begin{equation}  
   n
   \in
   \{1,2\}
   \mkern-3.3mu
   :
   \mkern4mu
   \text{\guillemotleft}
   \mkern1mu
   \Psi\mkern-3mu\left( x\right)
   \in
   \big\{
      c_n
      \phi_n\mkern-4mu\left( x\right)
      \mkern-2mu
      \big|
      \mkern2mu
      c_n
      \in
      \mathbb{C}
      ,
      \mkern2mu
      c_n
      \neq
      0
      \big.
   \big\}
   \mkern1mu
   \text{\guillemotright}
   \mkern-2mu
   =
   s_n
   \;\;\;\;  ,
\end{equation}
\smallskip

\noindent where $\phi_1\mkern-4mu\left( x\right)$ and $\phi_2\mkern-4mu\left( x\right)$ are spatially separated complex functions of $x$ localized at $x=0$ and $x=d$, in that order. The statement $s_n$ is true if $\Psi\mkern-3mu\left( x\right)$ is equal to the function $\phi_n\mkern-4mu\left( x\right)$ multiplied by some (non-zero) complex number $c_n$, otherwise $s_n$ is false.\\

\noindent Due to their spatial separation, the functions $\phi_1\mkern-4mu\left( x\right)$ and $\phi_2\mkern-4mu\left( x\right)$ can be made orthonormal over the interval $-\infty \le x \le +\infty$, to be exact,\smallskip

\begin{equation}  
   n,l
   \in
   \{1,2\}
   \mkern-3.3mu
   :
   \mkern2mu
   \langle
       \phi_n\mkern-4mu\left( x\mkern-2mu\right)
       \mkern-2mu|
       \phi_l\mkern-4mu\left( x\mkern-2mu\right)
   \mkern-2mu
   \rangle
   =
   \int_{-\infty}^{+\infty}
      \mkern-8mu
      \phi_n^\ast\mkern-4mu\left( x\right)
      \phi_l\mkern-4mu\left( x\right)
   \mkern-1mu
   \mathrm{d}x
   =
   \delta_{nl}
   \;\;\;\;  .
\end{equation}
\smallskip

\noindent As a result, the statements $s_1$ and $s_2$ cannot be true together, i.e.,\smallskip

\begin{equation}  
   n
   \in
   \{1,2\}
   ,
   \mkern1.5mu
   m
   =
   n-(-1)^n
   \mkern-3.3mu
   :
   \mkern7mu
   \left(
       s_n
      \sqcap
      s_m
   \right)
   \Leftrightarrow
   \bot
   \;\;\;\;  ,
\end{equation}
\smallskip

\noindent where the connective $\Leftrightarrow$ corresponds to the expression ``is equivalent to'' and $\bot$ stands for an arbitrary contradiction. In other words, the truth of $s_1$ means the falsity of $s_2$, and the truth of $s_2$ means the falsity of $s_1$.\\

\noindent However, it may be the case that $s_1$ and $s_2$ are false together, to be exact,\smallskip

\begin{equation}  
   \left(
       s_n
      \sqcup
      s_m
   \right)
   \nLeftrightarrow
   \top
   \;\;\;\;  ,
\end{equation}
\smallskip

\noindent where $\nLeftrightarrow$ stands for ``not equivalent to'' and $\top$ signifies an arbitrary tautology. For example, assume that the state of the quantum particle, just as it emerges from the double-slit plate, is described by a superposition of the functions $\phi_1\mkern-4mu\left( x\right)$ and $\phi_2\mkern-4mu\left( x\right)$, namely,\smallskip

\begin{equation}  
   \Psi\mkern-4mu\left( x\right)
   =
   c_1\phi_1\mkern-4mu\left( x\right)
   +
   c_2\phi_2\mkern-4mu\left( x\right)
   \;\;\;\;  .
\end{equation}
\smallskip

\noindent Then, $\Psi\mkern-0.5mu(\mkern-1mu x\mkern-1.5mu)$ is not an element of $\{c_1\phi_1\mkern-4mu\left(x\right)\mkern-2mu|\mkern2mu c_1\mkern-2mu\in\mkern-2mu\mathbb{C},c_1\neq0\}$, and neither it is an element of $\{c_2\phi_2\mkern-4mu\left(x\right)\mkern-2mu|\mkern2mu c_2\mkern-2mu\in\mkern-2mu\mathbb{C},c_2\neq0\}$, that is, both $s_1$ and $s_2$ are false.\\

\noindent Suppose that at the moment $t$ the particle reaches the screen behind the double-slit plate and at that time its state is given by\smallskip

\begin{equation} \label{SUP} 
   \Psi\mkern-4mu\left( x,t\right)
   =
   c_1\phi_1\mkern-4mu\left( x,t\right)
   +
   c_2\phi_2\mkern-4mu\left( x,t\right)
   \;\;\;\;  ,
\end{equation}
\smallskip

\noindent where $\langle \phi_n\mkern-4mu\left( x,t\right)\mkern-2mu|\phi_l\mkern-4mu\left( x,t\right) \rangle = \delta_{nl}$. Understanding that the function $\Psi\mkern-4mu\left( x,t\right)$ is orthonormal over the interval $-\infty\le{x}\le+\infty$, the following holds\smallskip

\begin{equation}  
   \langle
      \mkern-1mu
      \Psi\mkern-4mu\left( x,t\right)
      \mkern-2mu|
      \Psi\mkern-4mu\left( x,t\right)
      \mkern-1mu
   \rangle
   =
   \int_{-\infty}^{+\infty}
      \mkern-6mu
      \big|
         \mkern-2mu
         \Psi\mkern-4mu\left( x,t\right)
         \mkern-4mu
      \big|^2
   \mkern-1mu
   \mathrm{d}x
   =
   1
   \;\;\;\;  ,
\end{equation}
\smallskip

\noindent where, in accordance with (\ref{SUP}),\smallskip

\begin{equation}  
   \big|
      \mkern-1mu
      \Psi\mkern-4mu\left( x,t\right)
      \mkern-4mu
   \big|^2
   =
   \big|
      \mkern-0.5mu
      c_1\phi_1\mkern-4mu\left( x,t\right)
      \mkern-4mu
   \big|^2
   +
   \big|
      \mkern-0.5mu
      c_2\phi_2\mkern-4mu\left( x,t\right)
      \mkern-4mu
   \big|^2
   +
   c_1^\ast
   c_2
   \mkern1mu
   \phi_1^\ast\mkern-4mu\left( x,t\right)
   \mkern-1mu
   \phi_2\mkern-4mu\left( x\right)
   +
   c_2^\ast
   c_1
   \mkern1mu
   \phi_2^\ast\mkern-4mu\left( x,t\right)
   \mkern-1mu
   \phi_1\mkern-4mu\left( x,t\right)
   \;\;\;\;  .
\end{equation}
\smallskip

\noindent The cross terms in the above expression represent interference.\\

\noindent Those terms will drop off if a which-way detector is placed at the double-slit plate. To demonstrate this, assume that $|\mkern1mu{d_1}\mkern-1mu\rangle$ and $|\mkern1mu{d_2}\mkern-1mu\rangle$ are two possible quantum states of a which-way detector such that\smallskip

\begin{equation}  
   \langle
      \mkern-0.5mu 
      d_n\mkern-2mu|\mkern1mu{d_l}
      \mkern-1mu
   \rangle
   =
   \delta_{nl}
   \;\;\;\;  .
\end{equation}
\smallskip

\noindent In the presence of this detector, the quantum states of the paths $|\phi_1\rangle$ and $|\phi_2\rangle$ given by\smallskip

\begin{equation}  
   |\phi_n\rangle
   =
   \int_{-\infty}^{+\infty}
      \mkern-6mu
      \phi_{n}\mkern-4mu\left( x\right)
   \mkern-1mu
   |x\rangle
   \mathrm{d}x
   \approx
   \int_{x_n-\Delta{x}}^{x_n+\Delta{x}}
      \mkern-6mu
      \phi_{n}\mkern-4mu\left( x\right)
   \mkern-1mu
   |x\rangle
   \mathrm{d}x
   \;\;\;\;  ,
\end{equation}
\smallskip

\noindent where $x_1=0$, $x_2=d$ and $\Delta{x}$ is some positive value, get entangled with the corresponding states of the detector $|\mkern1mu{d_1}\mkern-1mu\rangle$ and $|\mkern1mu{d_2}\mkern-1mu\rangle$, so that immediately after the particle emerges from the double slit plate (equipped with the detector), the total quantum state of the experiment $|\Psi_{\text{exp}}\rangle$ is\smallskip

\begin{equation}  
   |\Psi_{\text{exp}}\rangle
   =
   \int_{-\infty}^{+\infty}
      \mkern-6mu
      \Psi_{\text{exp}}\mkern-4mu\left( x\right)
   \mkern-1mu
   |x\rangle
   \mathrm{d}x
   \;\;\;\;  ,
\end{equation}
\smallskip

\noindent where\smallskip

\begin{equation}  
   \Psi_{\text{exp}}\mkern-4mu\left( x\right)
   =
   c_1\phi_1\mkern-4mu\left( x\right)
   \mkern-1mu
   |\mkern1mu{d_1}\mkern-1mu\rangle
   +
   c_2\phi_2\mkern-4mu\left( x\right)
   \mkern-1mu
   |\mkern1mu{d_2}\mkern-1mu\rangle
   \;\;\;\;  .
\end{equation}
\smallskip

\noindent Due to orthogonality of $|\mkern1mu{d_1}\mkern-1mu\rangle$ and $|\mkern1mu{d_2}\mkern-1mu\rangle$, the cross terms will be missing in the square modulus of $\Psi_{\text{exp}}\mkern-4mu\left( x,t\right)$:\smallskip

\begin{equation}  
   \big|
      \mkern-1mu
      \Psi_{\text{exp}}\mkern-4mu\left( x,t\right)
      \mkern-4mu
   \big|^2
   =
   \big|
      \mkern-0.5mu
      c_1\phi_1\mkern-4mu\left( x,t\right)
      \mkern-4mu
   \big|^2
   +
   \big|
      \mkern-0.5mu
      c_2\phi_2\mkern-4mu\left( x,t\right)
      \mkern-4mu
   \big|^2
   \;\;\;\;  .
\end{equation}
\smallskip

\noindent Moreover, caused by the macroscopic nature of the which-way detector, the state $|\Psi_{\text{exp}}\rangle$ will evolve over some (short) period $\tau$ into one of the entangled states, meaning that the quantum particle will be reported by the macroscopic detector at exactly one slit. In symbols,\smallskip

\begin{equation}  
   c_1\phi_1\mkern-4mu\left( x\right)
   \mkern-1mu
   |\mkern1mu{d_1}\mkern-1mu\rangle
   +
   c_2\phi_2\mkern-4mu\left( x\right)
   \mkern-1mu
   |\mkern1mu{d_2}\mkern-1mu\rangle
   \stackrel{\tau}{\longrightarrow}
   \mkern6mu
   \text{\bf{either}}
   \mkern6mu
   c_1\phi_1\mkern-4mu\left( x,\tau\right)
   \mkern-1mu
   |\mkern1mu{d_1}\mkern-1mu\rangle
   \mkern6mu
   \text{\bf{or}}
   \mkern6mu
   c_2\phi_2\mkern-4mu\left( x,\tau\right)
   \mkern-1mu
   |\mkern1mu{d_2}\mkern-1mu\rangle
   \;\;\;\;  .
\end{equation}
\smallskip

\noindent \noindent Therefore, the atomic mathematical statements\smallskip

\begin{equation}  
   n
   \in
   \{1,2\}
   \mkern-3.3mu
   :
   \mkern4mu
   \text{\guillemotleft}
   \mkern1mu
   \Psi_{\text{exp}}\mkern-4mu\left( x,\tau\right)
   \in
   \big\{
      c_n
      \phi_n\mkern-4mu\left( x,\tau\right)
      \mkern-1mu
      |\mkern1mu{d_n}\mkern-1mu\rangle
      \big|
      \mkern2mu
      c_n
      \in
      \mathbb{C}
      ,
      \mkern2mu
      c_n
      \neq
      0
      \big.
   \big\}
   \mkern1mu
   \text{\guillemotright}
   \mkern-2mu
   =
   s_n^{\mkern2mu\prime}
   \;\;\;\;   
\end{equation}
\smallskip

\noindent will be neither true together nor false together, that is, $(s_1^{\mkern2mu\prime} \sqcap s_2^{\mkern2mu\prime})\Leftrightarrow\bot$ and $(s_1^{\mkern2mu\prime} \sqcup s_2^{\mkern2mu\prime})\Leftrightarrow\top$.\\

\noindent In contrast to propositional logic, assume that the notions of truth and falsity of a statement cannot be regarded as primitive; rather, a proof must be provided in order to accept that the statement is true or false. For a mathematical statement, such a proof can be either constructive or non-constructive. A non-constructive proof confirms the validity of a mathematical relation between expressions constituting a statement without providing an instance of those expressions. Whereas a constructive proof demonstrates that the mathematical relation is valid by creating an instance of the expressions. Take the mathematical statement such as \text{\guillemotleft}$a^b\in\mathbb{Q}$\text{\guillemotright}  where the symbols $a$ and $b$ denote certain irrational numbers: $a,b\in\mathbb{R}{\backslash}\mathbb{Q}$. To confirm the validity of the relation $\in$ between $a^b$ and the set of rational numbers, $\mathbb{Q}$, the non-constructive proof does not provide an instance of $a$ and $b$ but shows that the said relation is possible; whilst on the contrary, the constructive proof gives such an example: $(\sqrt{2})^{\log_{\sqrt{2}}3}=3$.\\

\noindent As to a proposition, its proof can be provided by the truth of the relating mathematical statements. Specifically, the truth of the mathematical statement $s_n$ (or $s_n^{\mkern2mu\prime}$) can be taken as \emph{positive evidence} witnessing the truth of the proposition $P_n$. By the same token, the truth of $s_m$ (or $s_m^{\mkern2mu\prime}$) can serve as \emph{negative evidence} demonstrating the falsity of $P_n$.\\

\noindent The problem is how to define a truth value of the proposition $P_n$ if neither evidence exists, that is, if both $s_n$ and $s_m$ are false and no which-way detector is present.\\

\noindent To state this problem formally, let us use the double-bracket notation $[\mkern-3.3mu[\cdot]\mkern-3.3mu]$ to express a truth value of a mathematical statement or a proposition.\\

\noindent Due to its nature, a mathematical statement – atomic and molecular alike – cannot be both true and false as well as neither true nor false. Hence, the relation between the set of mathematical statements and the set of truth values is a total surjective-only function, i.e.,\smallskip

\begin{equation}  
   v
   \mkern-3.3mu
   :
   \mkern2mu
   \mathbb{S}
   \to
   \mathbb{B}_{2}
   \;\;\;\;  ,
\end{equation}
\smallskip

\noindent where $\mathbb{S}$ denotes the set of mathematical statements and $\mathbb{B}_{2}$ stands for the set of two truth values, T (”true”) and F (”false”), which can be interpreted as integers 1 and 0, respectively. The image of a mathematical statement, for example, $s$, under this function can be denoted by $[\mkern-3.3mu[s]\mkern-3.3mu] = v(s)$.\\

\noindent Let us introduce a total bijective (i.e., both injective and surjective) function $b$ that takes truth values of two (different) mathematical statements to elements of $\mathbb{B}_{2}$:\smallskip

\begin{equation} \label{MAP} 
   b
   \mkern-3.3mu
   :
   \mkern2mu
   \mathbb{B}_{2}
   \times
   \mathbb{B}_{2}
   \to
   \mathbb{B}_{2}
   \;\;\;\;  .
\end{equation}
\smallskip

\noindent Using this function, the truth value of the proposition $P_n$ can be considered as the image of ordered pair $([\mkern-3.3mu[s_n]\mkern-3.3mu],[\mkern-3.3mu[s_m]\mkern-3.3mu])$ under $b$, that is,\smallskip

\begin{equation}  
   [\mkern-3.3mu[P_n]\mkern-3.3mu]
   =
   b
   \left(
      [\mkern-3.3mu[s_n]\mkern-3.3mu]
      ,
      [\mkern-3.3mu[s_m]\mkern-3.3mu]
   \right)
   \;\;\;\;  .
\end{equation}
\smallskip

\noindent Explicitly, the function $b$ will return 1 if $s_n$ is true, and $b$ will return 0 if $s_m$ is true; in symbols, $b(1,0)=1$ and $b(0,1)=0$.\\

\noindent Then again, both $s_n$ and $s_m$ can be false, so, there is one more pair, $(0,0)$. Given three different objects – i.e., three ordered pairs $(1,0)$, $(0,1)$ and $(0,0)$ – but only two elements of $\mathbb{B}_{2}$ to map them onto, one has a problem (which can be called \emph{the problem of an extra object}): What is the image of the pair $(0,0)$ under the function $b$? In other words, what truth value does the proposition $P_n$ have if $\Psi\mkern-4mu\left( x\right)$ is a superposition of the states $\phi_1\mkern-4mu\left( x\right)$ and $\phi_2\mkern-4mu\left( x\right)$? Symbolically, this problem can be presented as follows:\smallskip

\begin{equation} \label{PR1} 
   [\mkern-3.3mu[P_n]\mkern-3.3mu]
   =
   b
   \left(
      [\mkern-3.3mu[s_n]\mkern-3.3mu]
      ,
      [\mkern-3.3mu[s_m]\mkern-3.3mu]
   \right)
   =
   \left\{
      \begingroup\SmallColSep
      \begin{array}{r l}
         1
         &
         \mkern3mu
         ,
         \mkern12mu
         [\mkern-3.3mu[s_n]\mkern-3.3mu]
         =
         1
         \\[5pt]
         0
         &
         \mkern3mu
         ,
         \mkern12mu
         [\mkern-3.3mu[s_m]\mkern-3.3mu]
         =
         1
         \\[5pt]
         ?
         &
         \mkern3mu
         ,
         \mkern12mu
         [\mkern-3.3mu[s_n]\mkern-3.3mu]
         =
         [\mkern-3.3mu[s_m]\mkern-3.3mu]
         =
         0
      \end{array}
      \endgroup   
   \right.
   \;\;\;\;  .
\end{equation}
\smallskip

\section{Truth values of a propositional formula}  

\noindent The problem of an extra object also concerns an assignment of the truth values to a propositional formula (i.e., an expression involving finitely many logical connectives and propositions).\\

\noindent To ascertain this, let us turn to projection operators, i.e., self-adjoint operators with spectrum contained in the two-element set $\{0,1\}$. Such operators are in one-to-one correspondence with the closed linear subspaces of a Hilbert space $\mathcal{H}$ (i.e., a complex vector space upon which an inner or scalar product is defined) \cite{Mirsky}.\\

\noindent Let $\hat{P}$ be a projection operator; then, every unit vector $|\Psi\rangle\mkern-3mu\in\mkern-3mu\mathcal{H}$ can be decomposed uniquely as $|\Psi\rangle=|\psi\rangle+|\phi\rangle$ with $|\psi\rangle=\hat{P}|\Psi\rangle$ and $|\phi\rangle=\neg\hat{P}|\Psi\rangle$, where $\neg\hat{P}$ is the projection operator corresponding to the negation of $\hat{P}$ which can be expressed through the identical operator $\hat{1}$ and $\hat{P}$ as $\neg\hat{P}=\hat{1}-\hat{P}$, $|\psi\rangle$ belongs to the closed linear subspace $\mathcal{H}_{\hat{P}}$, while $|\phi\rangle$ lies in $\mathcal{H}_{\hat{P}}^{\perp}$, i.e., the closed linear subspace orthogonal to $\mathcal{H}_{\hat{P}}$. Therefore, $\hat{P}$ breaks the Hilbert space $\mathcal{H}$ into two orthogonal subspaces,\smallskip

\begin{equation}  
   \mathcal{H}
   =
   \mathcal{H}_{\hat{P}}
   \oplus
   \mathcal{H}_{\hat{P}}^{\perp}
   \;\;\;\;  ,
\end{equation}
\smallskip

\noindent such that $\hat{P}$ leaves any vector in $\mathcal{H}_{\hat{P}}$ invariant but annihilates any vector in $\mathcal{H}_{\hat{P}}^{\perp}$, namely,\smallskip

\begin{equation} 
   \mathcal{H}_{\hat{P}}
   =
   \left\{
      |\psi\rangle
      \in
      \mathcal{H}
      \textnormal{:}
      \mkern10mu
      \hat{P}
      |\psi\rangle
      =
      |\psi\rangle
      \mkern-2mu    
   \right\}
   \;\;\;\;  ,
\end{equation}
\\[-30pt]

\begin{equation}  
   \mathcal{H}_{\hat{P}}^{\perp}
   =
   \left\{
      |\phi\rangle
      \in
      \mathcal{H}
      \textnormal{:}
      \mkern10mu
      \hat{P}
      |\phi\rangle
      =
      0
      \mkern-2mu    
   \right\}
   \;\;\;\;  .
\end{equation}
\smallskip

\noindent For the Hermitian operator $A$ with a discrete orthonormal basis in each eigenspace (i.e., a subspace containing eigenvectors $|a_n\rangle$ of a given eigenvalue $a_n$), the projection operator $\hat{P}$ can be presented as\smallskip

\begin{equation}  
   \hat{P}_n
   =
   |a_n\rangle
   \langle{a_n}|
   \;\;\;\;  ,
\end{equation}
\smallskip

\noindent on condition that the eigenvalue $a_n$ is nondegenerate (that is, the eigenspace is 1-dimensional). Accordingly, two projection operators $\hat{P}_n$ and $\hat{P}_m$ of the same Hermitian operator $A$ satisfy\smallskip

\begin{equation}  
   \hat{P}_n
   \hat{P}_m
   =
   \hat{P}_m
   \hat{P}_n
   =
   \delta_{nm}
   \hat{P}_n
   \;\;\;\;  .
\end{equation}
\smallskip

\noindent In the case of the position operator $X$ (whose spectrum is continuous), the projection operators $\hat{P}_n$ are associated with the corresponding intervals $[x_n-\Delta{x},x_n+\Delta{x}]$, specifically,\smallskip

\begin{equation} \label{P} 
   \hat{P}_n
   =
   \int_{x_n-\Delta{x}}^{x_n+\Delta{x}}
      \mkern-4mu
      |x\rangle
      \langle{x}|
   \mkern2mu
   \mathrm{d}x
   \;\;\;\;  ,
\end{equation}
\smallskip

\noindent so that $\hat{P}_1$ and $\hat{P}_2$ are orthogonal if these intervals do not intersect (i.e., if $2\Delta{x}<d$).\\

\noindent The resolution of identity related to the given case is\smallskip

\begin{equation}  
   \int_{-\infty}^{+\infty}
      \mkern-4mu
      |x\rangle
      \langle{x}|
   \mkern2mu
   \mathrm{d}x
   =
   \hat{P}_1
   +
   \hat{P}_2
   +
   \sum_{k=1}^3
   \int_{A_k}^{B_k}
      \mkern-4mu
      |x\rangle
      \langle{x}|
   \mkern2mu
   \mathrm{d}x
   =
   1
   \;\;\;\;  ,
\end{equation}
\smallskip

\noindent where $A_k$ and $B_k$ are the endpoints of the intervals $I_k^{\text{plate}}$  containing $x$-coordinates situated on the side of the plate opposite to the source emitting quantum particles, i.e., behind the slits:\smallskip

\begin{equation}  
   I_k^{\text{plate}}
   =
   \left\{
      \begingroup\SmallColSep
      \begin{array}{l l}
         (-\infty \mkern2mu, \mkern3mu x_1-\Delta{x}]
         &
         \mkern3mu
         ,
         \mkern12mu
         k=1
         \\[5pt]
         [x_1+\Delta{x}\mkern2mu, \mkern3mux_2-\Delta{x}]
         &
         \mkern3mu
         ,
         \mkern12mu
         k=2
         \\[5pt]
         [x_2+\Delta{x} \mkern2mu, \mkern3mu +\infty)
         &
         \mkern3mu
         ,
         \mkern12mu
         k=3
      \end{array}
      \endgroup   
   \right.
   \;\;\;\;  .
\end{equation}
\smallskip

\noindent The eigenkets of the projection operators $\hat{P}_k^{\text{plate}}=\int_{A_k}^{B_k}\mkern-4mu|x\rangle\langle{x}|\mkern2mu\mathrm{d}x$ associated with these intervals, i.e., the vectors that meet the condition $\hat{P}_k^{\text{plate}}|\phi_k^{\text{plate}}\rangle=|\phi_k^{\text{plate}}\rangle$, can be written as\smallskip

\begin{equation}  
   |\phi_k^{\text{plate}}\rangle
   =
   \int_{A_k}^{B_k}
      \mkern-6mu
      \phi_k^{\text{plate}}\mkern-4mu\left( x\right)
   \mkern-1mu
   |x\rangle
   \mathrm{d}x
   \;\;\;\;  .
\end{equation}
\smallskip

\noindent Provided that the plate is impenetrable to the particles (and non-classical paths of the particles, which cause higher-order corrections to the interference pattern \cite{Sawant}, are absent), each function $\phi_k^{\text{plate}}\mkern-4mu\left( x\right)$ must be equal to zero in the interval $I_k^{\text{plate}}$ and so each eigenket $|\phi_k^{\text{plate}}\rangle$ must be the zero vector $|0\rangle$ (belonging to the zero subspace $\{0\}$). For this reason, every $\hat{P}_k^{\text{plate}}$ can be regarded as the zero operator $\hat{0}$ (that takes any vector to the zero vector). Accordingly, in this case the completeness relation has the form:\smallskip

\begin{equation}  
   \hat{P}_1
   +
   \hat{P}_2
   =
   \hat{1}
   \;\;\;\;  .
\end{equation}
\smallskip

\noindent Let $L(\mathcal{H})$ denote the set of the closed linear subspaces of the Hilbert space $\mathcal{H}$. As stated by \cite{Neumann}, the pair of elements in $L(\mathcal{H})$ that represent the distinct propositions $Q$ and $P$ are the subspaces $\mathcal{H}_{\hat{Q}}$ and $\mathcal{H}_{\hat{P}}$, respectively. These subspaces can be called \emph{related} or \emph{comparable} in case the mathematical statement $z={z_1}\sqcup{z_2}$ is true, that is, if the following holds\smallskip

\begin{equation}  
   [\mkern-3.3mu[z]\mkern-3.3mu]
   =
   \max
   \left\{
       [\mkern-3.3mu[z_1]\mkern-3.3mu]
      ,
       [\mkern-3.3mu[z_2]\mkern-3.3mu]
   \right\}
   =
   1
   \;\;\;\;  ,
\end{equation}
\smallskip

\noindent where $z_1$ and $z_2$ stand in for the mathematical statements that affirm the subset relation $\subseteq$ among $\mathcal{H}_{\hat{Q}}$ and $\mathcal{H}_{\hat{P}}$. Explicitly,\smallskip

\begin{equation}  
   \text{\guillemotleft}
   \mkern1mu
   \mathcal{H}_{\hat{Q}}
   \subseteq
   \mathcal{H}_{\hat{P}}
   \mkern1mu
   \text{\guillemotright}
   \mkern-2mu
   =
   z_1
   \;\;\;\;  ,
\end{equation}
\\[-36pt]

\begin{equation}  
   \text{\guillemotleft}
   \mkern1mu
   \mathcal{H}_{\hat{P}}
   \subseteq
   \mathcal{H}_{\hat{Q}}
   \mkern1mu
   \text{\guillemotright}
   \mkern-2mu
   =
   z_2
   \;\;\;\;  .
\end{equation}
\smallskip

\noindent Note that when $z$ is true, the projection operators $\hat{Q}$ and $\hat{P}$ commute (are compatible), explicitly, $\hat{Q}\hat{P}=\hat{P}\hat{Q}$.\\

\noindent Contrastively, the subspaces $\mathcal{H}_{\hat{Q}}$ and $\mathcal{H}_{\hat{P}}$ can be called \emph{orthogonal} if the mathematical statement $w={w_1}\sqcup{w_2}$ is true, i.e., if\smallskip

\begin{equation}  
   [\mkern-3.3mu[w]\mkern-3.3mu]
   =
   \max
   \left\{
       [\mkern-3.3mu[w_1]\mkern-3.3mu]
      ,
       [\mkern-3.3mu[w_2]\mkern-3.3mu]
   \right\}
   =
   1
   \;\;\;\;  ,
\end{equation}
\smallskip

\noindent where $w_1$ and $w_2$ substitute for the mathematical statements asserting the orthogonality relation among $\mathcal{H}_{\hat{Q}}$ and $\mathcal{H}_{\hat{P}}$:\smallskip

\begin{equation}  
   \text{\guillemotleft}
   \mkern1mu
   \mathcal{H}_{\hat{Q}}
   \subseteq
   \mathcal{H}_{\hat{P}}^{\perp}
   \mkern1mu
   \text{\guillemotright}
   \mkern-2mu
   =
   w_1
   \;\;\;\;  ,
\end{equation}
\\[-32pt]

\begin{equation}  
   \text{\guillemotleft}
   \mkern1mu
   \mathcal{H}_{\hat{P}}
   \subseteq
   \mathcal{H}_{\hat{Q}}^{\perp}
   \mkern1mu
   \text{\guillemotright}
   \mkern-2mu
   =
   w_2
   \;\;\;\;  .
\end{equation}
\smallskip

\noindent Note that in case $w$ is true, the projection operators $\hat{Q}$ and $\hat{P}$ are compatible and orthogonal, i.e., $\hat{Q}\hat{P}=\hat{P}\hat{Q}=0$.\\

\noindent Clearly, if the subspaces $\mathcal{H}_{\hat{Q}}$ and $\mathcal{H}_{\hat{P}}$ are orthogonal, they are incomparable, and if they are comparable, they are not orthogonal. To be exact, except for the subspaces $\mathcal{H}_{\hat{0}}=\{0\}$ and $\mathcal{H}_{\hat{1}}=\mathcal{H}$, the statements $w$ and $z$ cannot be true together:\smallskip

\begin{equation}  
   \left(
       z
      \sqcap
      w
   \right)
   \Leftrightarrow
   \bot
   \;\;\;\;  .
\end{equation}
\smallskip

\noindent Even so, $z$ and $w$ may be false together:\smallskip

\begin{equation}  
   \left(
      z
      \sqcup
      w
   \right)
   \nLeftrightarrow
   \top
   \;\;\;\;  .
\end{equation}
\smallskip

\noindent In particular, if the projection operators $\hat{Q}$ and $\hat{P}$ are not compatible, i.e., $\hat{Q}\hat{P}\neq\hat{P}\hat{Q}$, then $\mathcal{H}_{\hat{Q}}\nsubseteq\mathcal{H}_{\hat{P}}$ and $\mathcal{H}_{\hat{P}}\nsubseteq\mathcal{H}_{\hat{Q}}$, as well as $\mathcal{H}_{\hat{Q}}\nsubseteq\mathcal{H}_{\hat{P}}^{\perp}$ and $\mathcal{H}_{\hat{P}}\nsubseteq\mathcal{H}_{\hat{Q}}^{\perp}$.\\

\noindent Let the inequality $Q{\mkern2mu\le\mkern2mu}P$ be expressible as $Q{\mkern2mu\sqcap\mkern2mu}P{\mkern4mu\Leftrightarrow\mkern2mu}Q$ or as $Q{\mkern2mu\sqcup\mkern2mu}P{\mkern2mu\Leftrightarrow\mkern2mu}P$. Also, let the inequality $Q{\mkern2mu\ge\mkern2mu}P$ be expressible as $Q{\mkern2mu\sqcap\mkern2mu}P{\mkern2mu\Leftrightarrow\mkern2mu}P$ or as $Q{\mkern2mu\sqcup\mkern2mu}P{\mkern2mu\Leftrightarrow\mkern2mu}Q$. Then, one can construct the propositional formula $Q{\mkern2mu\lesseqgtr\mkern2mu}P$ meaning\smallskip

\begin{equation}  
   Q{\mkern2mu\lesseqgtr\mkern2mu}P
   \Leftrightarrow
   \left(
      Q{\mkern2mu\le\mkern2mu}P
   \right)
   \sqcup
   \left(
      Q{\mkern2mu\ge\mkern2mu}P
   \right)
   \;\;\;\;  .
\end{equation}
\smallskip

\noindent It can be phrased as the following statement: {\guillemotleft}Two distinct propositions about the same quantum system can be true together{\guillemotright}. Given that for the above formula the equivalences $[\mkern-3.3mu[z]\mkern-3.3mu]=1$ and $[\mkern-3.3mu[w]\mkern-3.3mu]=1$ act respectively as positive and negative evidence, its truth value can be considered as the image of the ordered pair $([\mkern-3.3mu[z]\mkern-3.3mu],[\mkern-3.3mu[w]\mkern-3.3mu])$ under the map (\ref{MAP}), i.e.,\smallskip

\begin{equation}  
   [\mkern-3.3mu[Q{\mkern2mu\lesseqgtr\mkern2mu}P]\mkern-3.3mu]
   =
   b
   \left(
      [\mkern-3.3mu[z]\mkern-3.3mu]
      ,
      [\mkern-3.3mu[w]\mkern-3.3mu]
   \right)
   \;\;\;\;  .
\end{equation}
\smallskip

\noindent Specifically, $Q$ and $P$ can be true together if $\mathcal{H}_{\hat{Q}}$ and $\mathcal{H}_{\hat{P}}$ are comparable and, hence, not orthogonal; on the contrary, $Q$ and $P$ cannot be true together if $\mathcal{H}_{\hat{Q}}$ and $\mathcal{H}_{\hat{P}}$ are orthogonal, therefore, incomparable. That is, $[\mkern-3.3mu[Q{\mkern2mu\lesseqgtr\mkern2mu}P]\mkern-3.3mu]$ is equal to 1 on the pair $(1,0)$ and is equal to 0 on the pair $(0,1)$.\\

\noindent Again, similar to the situation with the atomic proposition $P_n$, the third pair exists, $(0,0)$, which gives rise to the problem of an extra object. The problem is this: How should a truth value of $Q{\mkern2mu\lesseqgtr\mkern2mu}P$ be defined in case that neither evidence exists for this propositional formula, that is, if the subspaces $\mathcal{H}_{\hat{Q}}$ and $\mathcal{H}_{\hat{P}}$ representing $Q$ and $P$ are neither comparable nor orthogonal? Symbolically,\smallskip

\begin{equation} \label{PR2} 
   [\mkern-3.3mu[Q{\mkern2mu\lesseqgtr\mkern2mu}P]\mkern-3.3mu]
   =
   b
   \left(
      [\mkern-3.3mu[z]\mkern-3.3mu]
      ,
      [\mkern-3.3mu[w]\mkern-3.3mu]
   \right)
   =
   \left\{
      \begingroup\SmallColSep
      \begin{array}{r l}
         1
         &
         \mkern3mu
         ,
         \mkern12mu
         [\mkern-3.3mu[z]\mkern-3.3mu]
         =
         1
         \\[5pt]
         0
         &
         \mkern3mu
         ,
         \mkern12mu
         [\mkern-3.3mu[w]\mkern-3.3mu]
         =
         1
         \\[5pt]
         ?
         &
         \mkern3mu
         ,
         \mkern12mu
         [\mkern-3.3mu[z]\mkern-3.3mu]
         =
         [\mkern-3.3mu[w]\mkern-3.3mu]
         =
         0
      \end{array}
      \endgroup   
   \right.
   \;\;\;\;  .
\end{equation}
\smallskip

\section{Three-valued semantics}  

\noindent An apparently justified solution to the problem of an extra object is to abandon the principle, according to which the set of the truth values must be limited to only two elements. Particularly, one can suggest adding a third element to the set $\{\text{T},\text{F}\}$, for example, U (i.e., the ``undefined'' or ``undetermined'' truth value, as it is assumed in Kleene's ``(strong) logic of indeterminacy'' \cite{Kleene} and Priest's ``logic of paradox'' \cite{Priest}, respectively). This element can be interpreted as a real number lying between 0 and 1, say, $\rfrac{1}{2}$, in agreement with {\L}ukasiewicz logic $\mathrm{L}_3$ \cite{Lukasiewicz}. The nature of the added element is supposed to be the same as the one of the truth values T and F (otherwise, it would be hard to guarantee the consistency and unambiguity of what constitutes and what does not constitute a set of the truth values \cite{Gottwald}).\\

\noindent As a result, the simple resolution of the extra object problem immediately follows. Indeed, using the total bijective function\smallskip

\begin{equation} 
   b
   \mkern-3.3mu
   :
   \mkern2mu
   \mathbb{B}_{2}
   \times
   \mathbb{B}_{2}
   \to
   \mathbb{B}_{3}
   \;\;\;\;   
\end{equation}
\smallskip

\noindent whose codomain $\mathbb{B}_{3}$ is the collection of three elements, $\{\text{T},\text{U},\text{F}\}$ or $\{1,\rfrac{1}{2},0\}$, a truth value of the proposition $P_n$ can be defined by\smallskip

\begin{equation} \label{MVL} 
   [\mkern-3.3mu[P_n]\mkern-3.3mu]
   =
   b
   \left(
      [\mkern-3.3mu[s_n]\mkern-3.3mu]
      ,
      [\mkern-3.3mu[s_m]\mkern-3.3mu]
   \right)
   =
   \left\{
      \begingroup\SmallColSep
      \begin{array}{r l}
         1
         &
         \mkern3mu
         ,
         \mkern12mu
         [\mkern-3.3mu[s_n]\mkern-3.3mu]
         =
         1
         \\[5pt]
         0
         &
         \mkern3mu
         ,
         \mkern12mu
         [\mkern-3.3mu[s_m]\mkern-3.3mu]
         =
         1
         \\[5pt]
         \rfrac{1}{2}
         &
         \mkern3mu
         ,
         \mkern12mu
         [\mkern-3.3mu[s_n]\mkern-3.3mu]
         =
         [\mkern-3.3mu[s_m]\mkern-3.3mu]
         =
         0
      \end{array}
      \endgroup   
   \right.
   \;\;\;\;  .
\end{equation}
\smallskip

\noindent Attractive as this proposal (initially introduced by Reichenbach \cite{Reichenbach}) might seem, it is open to serious objection.\\

\noindent First, why are there two different sets of the truth values – that is, the set $\{\text{T},\text{F}\}$ for statements of mathematical relations (i.e., analytic statements) and the set $\{\text{T},\text{U},\text{F}\}$ for propositions (i.e., synthetic statements)? Since the existence of a precisely cut distinction between analytic and synthetic statements is doubtful \cite{Rey}, the presence of two sets of the truth values appears unlikely.\\

\noindent Second, if the intermediate truth value U is assigned to at least some proposition(s) in a propositional formula, how is one to determine a truth value thereof? It is known to be a major problem for any multi-valued semantics since the act of ascertaining such a value is arbitrary by its very nature \cite{Isham}. For example, let us take the truth function of conjunction, $\sqcap$, and suppose that its operands have the truth values U and F. Then, according to Kleene's ``(strong) logic of indeterminacy'' \cite{Humberstone}, the said function must return F. However, in accordance with Bochvar's ``internal'' three-valued logic \cite{Bergmann}, the same function must produce U.\\

\noindent Third, let’s assume that if both slits are open and so the truth value U is assigned to both $P_1$ and $P_2$ in accordance with (\ref{MVL}), the propositional formulas ${P_1}\sqcup{P_2}$ and ${P_1}\sqcap{P_2}$ have equal truth values, namely, U. Now, suppose that at some time, a which-way detector is placed at the double-slit plate. After that, ${P_1}\sqcup{P_2}$ and ${P_1}\sqcap{P_2}$ transform respectively into a tautology and a contradiction. Hence, unlike the endpoint truth values T and F, the intermediate truth value U gets destroyed as soon as the observation is made. Recalling that U belongs to the same class of the objects as the truth values T and F do, and so it must survive the observation like T and F do, one comes to absurdity.\enlargethispage{\baselineskip}\enlargethispage{\baselineskip}\\

\noindent To avoid the above arguments, the problem of an extra object must be solved using a semantics in which propositions may only have two possible truth values, T and F.\\

\noindent Such solutions will be examined in the next sections of the paper.\\

\section{Birkhoff and von Neumann’s proposal}  

\noindent A solution proposed by Birkhoff and von Neumann \cite{Birkhoff, Piron} is to assume that the function (\ref{MAP}) is a total surjective but not injective function. This function is not injective because it associates two elements of $\mathbb{B}_{2}\times\mathbb{B}_{2}$ with one and the same element of $\mathbb{B}_{2}$. As a result, the image of the pair $(0,0)$ under $b$ may be equal to either 1 together with the pair $(1,0)$, or 0 together with the pair $(0,1)$. In either case,\smallskip

\begin{equation} \label{ABN} 
   b(0,0)
   \in
   \mathbb{B}_{2}
   \;\;\;\;  .
\end{equation}
\smallskip

\noindent On the other hand, using truth tables it is straightforward to demonstrate that for any two mathematical statements, say, $r_1$ and $r_2$, such that $({r_1}\mkern2mu\sqcap\mkern2mu{r_2})\mkern-4mu\Leftrightarrow\mkern-4mu\bot$ and $({r_1}\mkern2mu\sqcup\mkern2mu{r_2})\mkern-4mu\nLeftrightarrow\mkern-4mu\top$, the logical biconditional holds:\smallskip

\begin{equation}  
   r_n
   \sqcup
   \left(
      \neg{r_1}
      \sqcap
      \neg{r_2}
   \right)
   \Leftrightarrow
   \neg{r_m}
   \;\;\;\;  .
\end{equation}
\smallskip

\noindent Hence, the application of the proposal (\ref{ABN}) to (\ref{PR1}) makes a truth value of the atomic proposition $P_n$ subject to a truth value of either the statement $s_n$ or the statement $s_m$ alone, but not both.\\

\noindent Concretely, in case $b(0,0)=1$, a truth value of $P_n$ is determined by\smallskip

\begin{equation} \label{VALP} 
   [\mkern-3.3mu[P_n]\mkern-3.3mu]
   =
   1
   -
   [\mkern-3.3mu[s_{n-(-1)^n}]\mkern-3.3mu]
   \;\;\;\;  .
\end{equation}
\smallskip

\noindent In contrast, if $b(0,0)=0$, then the assignment of a truth value to $P_n$ is given by\smallskip

\begin{equation}  
   [\mkern-3.3mu[P_n]\mkern-3.3mu]
   =
   [\mkern-3.3mu[s_n]\mkern-3.3mu]
   \;\;\;\;  .
\end{equation}
\smallskip

\noindent In both cases, if either $s_1$ or $s_2$ is true, then $[\mkern-3.3mu[P_n]\mkern-3.3mu]=1$ while $[\mkern-3.3mu[P_m]\mkern-3.3mu]=0$. Providing truth values of the propositional formulas ${P_1}\sqcap{P_2}$ and ${P_1}\sqcup{P_2}$ are determined by truth values of its components, namely,\newline

\begin{equation}  
   [\mkern-3.3mu[{P_1}\sqcap{P_2}]\mkern-3.3mu]
   =
   \min
   \left\{
       [\mkern-3.3mu[P_1]\mkern-3.3mu]
      ,
       [\mkern-3.3mu[P_2]\mkern-3.3mu]
   \right\}
   \;\;\;\;  ,
\end{equation}
\\[-36pt]

\begin{equation}  
   [\mkern-3.3mu[{P_1}\sqcup{P_2}]\mkern-3.3mu]
   =
   \max
   \left\{
       [\mkern-3.3mu[P_1]\mkern-3.3mu]
      ,
       [\mkern-3.3mu[P_2]\mkern-3.3mu]
   \right\}
   \;\;\;\;  ,
\end{equation}
\smallskip

\noindent one finds that $[\mkern-3.3mu[{P_1}\sqcap{P_2}]\mkern-3.3mu]=0$ but $[\mkern-3.3mu[{P_1}\sqcup{P_2}]\mkern-3.3mu]=1$. That is, if either evidence for $P_n$ is present (e.g., either slit is open, or a which-way detector reports the quantum particle at either slit), the complex statement {\guillemotleft}$P_1$ or $P_2$, but not, $P_1$ and $P_2${\guillemotright} written down as exclusive disjunction ${P_1}\mkern2mu\underline{\sqcup}\mkern2mu{P_2}$, namely,\smallskip

\begin{equation}  
   P_1
   \mkern2mu
   \underline{\sqcup}
   \mkern2mu
   P_2
   \mkern4mu
   \Leftrightarrow
   \mkern4mu
   \left(
      {P_1}\sqcup{P_2}
   \right)
   \sqcap
   \neg
   \left(
      {P_1}\sqcap{P_2}
   \right)
   \;\;\;\;  ,
\end{equation}
\smallskip

\noindent is a tautology. The truthfulness of the statement ${P_1}\mkern2mu\underline{\sqcup}\mkern2mu{P_2}$ suggests that the quantum particle behaves as a classically defined particle.\\

\noindent Now, consider the case where both $s_1$ and $s_2$ are false (that is, the case where state of the quantum particle at the instant it passes the double-slit plate is described by a superposition). Assume that $b(0,0)=0$. This assumption brings on the equivalence $[\mkern-3.3mu[P_1]\mkern-3.3mu]=[\mkern-3.3mu[P_2]\mkern-3.3mu]=0$, which means that before being recorded on the screen the quantum particle went through neither slit. This does not seem to make much physical sense. Hence, so long as $b$ is a total non-injective surjective function, the image of the pair $(0,0)$ under this function should be 1. In symbols,\smallskip

\begin{equation}  
   \begingroup
   \begin{array}{r c c c}
      b:
      &
      \mathbb{B}_2\times\mathbb{B}_2
      &
      \to
      &
      \mathbb{B}_2
      \\[5pt]
      \hfill
      &
      (0,0)
      &
      \mapsto
      &
      1
   \end{array}
   \endgroup
   \;\;\;\;  ,
\end{equation}
\smallskip

\noindent where the second part is read: “$(0,0)$ maps onto 1”.\\

\noindent Applying the above to (\ref{PR2}) results in the following assignment:\smallskip

\begin{equation}  
   [\mkern-3.3mu[Q{\mkern2mu\lesseqgtr\mkern2mu}P]\mkern-3.3mu]
   =
   1
   -
   [\mkern-3.3mu[w]\mkern-3.3mu]
   \;\;\;\;  .
\end{equation}
\smallskip

\noindent Since the mathematical statement $w$ is either true or false, any two distinct propositions about the same quantum system are either able or unable to be true together. This fact implies that every pair of elements in $L(\mathcal{H})$ is either ordered or not ordered by the subset relation $\subseteq$. In this way, $L(\mathcal{H})$ proves to be a set with a partial order (a poset).\\

\noindent Being elements of the poset, any two subspaces in $L(\mathcal{H})$, say $\mathcal{H}_{\hat{Q}}$ and $\mathcal{H}_{\hat{P}}$, may have a meet (denoted $\mathcal{H}_{\hat{Q}}\wedge\mathcal{H}_{\hat{P}}$) and a join (denoted $\mathcal{H}_{\hat{Q}}\vee\mathcal{H}_{\hat{P}}$), regardless of compatibility between the projection operators $\hat{Q}$ and $\hat{P}$ that correspond to those subspaces. Particularly, since for any set of subsets, the set-intersection $\cap$ interprets meet $\wedge$, the meet operation on elements of the poset $L(\mathcal{H})$ can be defined as follows\smallskip

\begin{equation}  
   \mathcal{H}_{\hat{Q}}
   \wedge
   \mathcal{H}_{\hat{P}}
   =
   \mathcal{H}_{\hat{Q}}
   \cap
   \mathcal{H}_{\hat{P}}
\;\;\;\;  .
\end{equation}
\smallskip

\noindent Furthermore, because the subspace $\mathcal{H}_{\neg\hat{Q}}$ is the set of all vectors of $\mathcal{H}$ that are not in $\mathcal{H}_{\hat{Q}}$, except the zero subspace, $\{0\}$, namely,\smallskip

\begin{equation} 
   \mathcal{H}_{\neg\hat{Q}}
   =
   \left\{
      |\psi\rangle
      \in
      \mathcal{H}
      \textnormal{:}
      \mkern7mu
       \left(
         \hat{1}
         -
         \hat{Q}
       \right)
      \mkern-2mu    
      |\psi\rangle
      =
      |\psi\rangle
      \mkern-2mu    
   \right\}
   \;\;\;\;  ,
\end{equation}
\smallskip

\noindent it holds $\mathcal{H}_{\neg\hat{Q}}=\mathcal{H}_{\hat{Q}}^{\perp}$. Similarly, $\mathcal{H}_{\neg\hat{P}}=\mathcal{H}_{\hat{P}}^{\perp}$. Thus, the join $\mathcal{H}_{\hat{Q}}\vee\mathcal{H}_{\hat{P}}$ can be derived from the meet operation using De Morgan's law \cite{Halmos}, i.e.,\smallskip

\begin{equation}  
   \mathcal{H}_{\hat{Q}}
   \vee
   \mathcal{H}_{\hat{P}}
   =
   \left(
      \mathcal{H}_{\neg\hat{Q}}
      \cap
      \mathcal{H}_{\neg\hat{P}}
   \right)^{\perp}
\;\;\;\;  .
\end{equation}
\smallskip

\noindent In those circumstances, the poset $L(\mathcal{H})$ can be held as a complete lattice (usually called \emph{a Hilbert lattice}).\\

\noindent Provided that ${Q}\sqcap{P}$ and ${Q}\sqcup{P}$ are represented by the meet and join of the subspaces $\mathcal{H}_{\hat{Q}}$ and $\mathcal{H}_{\hat{P}}$, truth values of these propositional formulas are supposed to be assigned after the valuation (\ref{VALP}). Specifically,\smallskip

\begin{equation}  
   [\mkern-3.3mu[{Q}\sqcap{P}]\mkern-3.3mu]
   =
   1
   -
   [\mkern-3.3mu[
      \text{\guillemotleft}
      |\Psi\rangle
      \in
      \left(
      \mathcal{H}_{\hat{Q}}\wedge\mathcal{H}_{\hat{P}}
      \right)^{\perp}
      \mkern-3mu
      \text{\guillemotright}
   ]\mkern-3.3mu]
   \;\;\;\;  ,
\end{equation}
\\[-36pt]

\begin{equation}  
   [\mkern-3.3mu[{Q}\sqcup{P}]\mkern-3.3mu]
   =
   1
   -
   [\mkern-3.3mu[
      \text{\guillemotleft}
      |\Psi\rangle
      \in
      \left(
      \mathcal{H}_{\hat{Q}}\vee\mathcal{H}_{\hat{P}}
      \right)^{\perp}
      \mkern-3mu
      \text{\guillemotright}
   ]\mkern-3.3mu]
   \;\;\;\;  ,
\end{equation}
\smallskip

\noindent where $|\Psi\rangle$ is the vector of $\mathcal{H}$ describing the state of the quantum particle when it comes out from the double-slit plate:\smallskip

\begin{equation}  
   |\Psi\rangle
   =
   \int
      \mkern-6mu
      \Psi\mkern-4mu\left( x\right)
   \mkern-1mu
   |x\rangle
   \mathrm{d}x
   \;\;\;\;  .
\end{equation}
\smallskip

\noindent If $\Psi\mkern-4mu\left( x\right)$ is a superposition $c_1\phi_1\mkern-4mu\left( x\right)+c_2\phi_2\mkern-4mu\left( x\right)$, then $|\Psi\rangle$ is a sum of vectors\smallskip

\begin{equation}  
   |\Psi\rangle
   =
   c_1\mkern-0.5mu|\phi_1\rangle
   +
   c_2\mkern-0.5mu|\phi_1\rangle
   \;\;\;\;   
\end{equation}
\smallskip

\noindent whose constituents are in the subspaces $\mathcal{H}_{\hat{P}_n}$ corresponding to the projection operators $\hat{P}_n$:\smallskip

\begin{equation} 
   \mathcal{H}_{\hat{P}_n}
   =
   \left\{
      |\phi_n\rangle
      \in
      \mathcal{H}
      \textnormal{:}
      \mkern10mu
      \hat{P}_n
      |\phi_n\rangle
      =
      |\phi_n\rangle
      \mkern-2mu    
   \right\}
   \;\;\;\;  .
\end{equation}
\smallskip

\noindent In that case, the mathematical statement {\guillemotleft}$\left(c_1\mkern-0.5mu|\phi_1\rangle +c_2\mkern-0.5mu|\phi_1\rangle \right)\mkern-1.5mu\in\mkern-1.5mu\{c_{n-(-1)^n}|\phi_{n-(-1)^n}\rangle\}${\guillemotright} is false, and so – in accordance with (\ref{VALP}) – one gets\smallskip

\begin{equation}  
   [\mkern-3.3mu[P_n]\mkern-3.3mu]
   =
   1
   -
   [\mkern-3.3mu[
      \text{\guillemotleft}
      \mkern-2mu
      \left(c_1\mkern-0.5mu|\phi_1\rangle +c_2\mkern-0.5mu|\phi_2\rangle \right)\mkern-1.5mu\in\mkern-1.5mu\{c_{n-(-1)^n}|\phi_{n-(-1)^n}\mkern-1.5mu\rangle\}
      \text{\guillemotright}
   ]\mkern-3.3mu]
   =
   1
   \;\;\;\;  .
\end{equation}
\smallskip

\noindent Recalling that the propositional formulas ${P_1}\sqcap{P_2}$ and ${P_1}\sqcup{P_2}$ are truth-functional, this entails $[\mkern-3.3mu[{P_1}\sqcap{P_2}]\mkern-3.3mu]=[\mkern-3.3mu[{P_1}\sqcup{P_2}]\mkern-3.3mu]=1$. That is, if two slits are open (and no which-way detector is present), the statement ${P_1}\mkern2mu\underline{\sqcup}\mkern2mu{P_2}$ whose truth value is defined by\smallskip

\begin{equation}  
   [\mkern-3.3mu[{P_1}\mkern2mu\underline{\sqcup}\mkern2mu{P_2}]\mkern-3.3mu]
   =
   \min
   \left\{
       [\mkern-3.3mu[{P_1}\mkern2mu{\sqcup}\mkern2mu{P_2}]\mkern-3.3mu]
       ,
      1
      - 
      [\mkern-3.3mu[{P_1}\mkern2mu{\sqcap}\mkern2mu{P_2}]\mkern-3.3mu]
   \right\}
   \;\;\;\;   
\end{equation}
\smallskip

\noindent is a contradiction.\\

\noindent However, this may not suggest that the quantum particle is wave-like. The falsity of ${P_1}\mkern2mu\underline{\sqcup}\mkern2mu{P_2}$ can be explained away by non-distributivity of the logical operators $\sqcap$ and $\sqcup$ over each other caused by nondistributiveness of the lattice $L(\mathcal{H})$. To demonstrate this, let us suppose that the vector $|\Psi\rangle=c_1\mkern-0.5mu|\phi_1\rangle+c_2\mkern-0.5mu|\phi_1\rangle$ belongs to the linear subspace $\mathcal{H}_{\hat{Q}}$, which is in one-one correspondence with the projection operator $\hat{Q}$ representing some proposition $Q$.\\

\noindent Consider the propositional formula $\neg{Q}\sqcup({P_1}\sqcap{P_2})$, where $\neg{Q}$ is the negation of $Q$. Because $\neg{Q}$ is false in the state $|\Psi\rangle\mkern-3mu\in\mkern-3mu\mathcal{H}_{\hat{Q}}$, the truth value of the said formula therein can be defined as\smallskip

\begin{equation}  
   [\mkern-3.3mu[\neg{Q}\sqcup({P_1}\sqcap{P_2})]\mkern-3.3mu]
   =
   \max\mkern-2mu\big\{
      [\mkern-3.3mu[\neg{Q}]\mkern-3.3mu]
      ,
      [\mkern-3.3mu[{P_1}\sqcap{P_2}]\mkern-3.3mu]
   \big\}
   =
   [\mkern-3.3mu[{P_1}\sqcap{P_2}]\mkern-3.3mu]
   \;\;\;\;  .
\end{equation}
\smallskip

\noindent Since $\mathcal{H}_{\hat{P}_1}\wedge\mathcal{H}_{\hat{P}_2}$ is the zero subspace, $\{0\}$, orthogonal to the identical subspace, $\mathcal{H}$, the mathematical statement {\guillemotleft}$|\Psi\rangle\in(\mathcal{H}_{\hat{P}_1}\wedge\mathcal{H}_{\hat{P}_2})^{\perp}\mkern-0.5mu${\guillemotright} is true for any vector $|\Psi\rangle$ in $\mathcal{H}$. Consequently,\smallskip

\begin{equation}  
   [\mkern-3.3mu[{P_1}\sqcap{P_2}]\mkern-3.3mu]
   =
   1
   -
   [\mkern-3.3mu[
      \text{\guillemotleft}
      |\Psi\rangle
      \in
      (
      \mathcal{H}_{\hat{P}_1}\wedge\mathcal{H}_{\hat{P}_2}
      )^{\perp}
      \mkern-0.5mu
      \text{\guillemotright}
   ]\mkern-3.3mu]
   =
   0
   \;\;\;\;  .
\end{equation}
\smallskip

\noindent By the same token, as the mathematical statement {\guillemotleft}$|\Psi\rangle\mkern-2mu\in\mkern-2mu(\mathcal{H}_{\hat{P}_1}\vee\mathcal{H}_{\hat{P}_2})^{\perp}\mkern-0.5mu${\guillemotright} equivalent to {\guillemotleft}$|\Psi\rangle\mkern-2mu\in\mkern-2mu\{0\}${\guillemotright} is false for any nonzero vector $|\Psi\rangle$ in $\mathcal{H}$, one gets\smallskip

\begin{equation}  
   [\mkern-3.3mu[{P_1}\sqcup{P_2}]\mkern-3.3mu]
   =
   1
   -
   [\mkern-3.3mu[
      \text{\guillemotleft}
      |\Psi\rangle
      \in
      (
      \mathcal{H}_{\hat{P}_1}\vee\mathcal{H}_{\hat{P}_2}
      )^{\perp}
      \mkern-0.5mu
      \text{\guillemotright}
   ]\mkern-3.3mu]
   =
   1
   \;\;\;\;  .
\end{equation}
\smallskip

\noindent Now, take the propositional formula $(\neg{Q}\sqcup{P_1})\sqcap(\neg{Q}\sqcup{P_2})$. In the state $(c_1\mkern-0.5mu|\phi_1\rangle+c_2\mkern-0.5mu|\phi_1\rangle)\in\mkern-2mu\mathcal{H}_{\hat{Q}}$, the truth value of this formula is defined by\smallskip

\begin{equation} \label{PART2} 
   [\mkern-3.3mu[(\neg{Q}\sqcup{P_1})\sqcap(\neg{Q}\sqcup{P_2})]\mkern-3.3mu]
   =
   \min\mkern-2mu\big\{
      [\mkern-3.3mu[\neg{Q}\sqcup{P_1}]\mkern-3.3mu]
      ,
      [\mkern-3.3mu[\neg{Q}\sqcup{P_2}]\mkern-3.3mu]
   \big\}
   =
   \min\mkern-2mu\big\{
      [\mkern-3.3mu[P_1]\mkern-3.3mu]
      ,
      [\mkern-3.3mu[P_2]\mkern-3.3mu]
   \big\}
   \;\;\;\;  .
\end{equation}
\smallskip

\noindent Then again, taking into consideration that the subspace $\mathcal{H}_{\hat{Q}}$ intersects (meets) the subspace $\mathcal{H}_{\hat{P}_m}$ at $\{0\}$, that is,\smallskip

\begin{equation} 
   \mathcal{H}_{\hat{Q}}
   \cap
   \mathcal{H}_{\hat{P}_m}
   =
   \left\{
      |\psi\rangle
      \mkern-3mu
      \in
      \mkern-3mu
      \mathcal{H}
      \textnormal{:}
      \mkern10mu
      \hat{Q}
      \mkern-0.5mu    
      |\psi\rangle
      =
      |\psi\rangle
      \mkern6mu      
      \textnormal{\bf{and}}
      \mkern6mu      
      \hat{P}_m
      \mkern-2mu    
      |\psi\rangle
      =
      |\psi\rangle
      \mkern-2mu    
   \right\}
   =
   \{0\}
   \;\;\;\;  ,
\end{equation}
\smallskip

\noindent the statement {\guillemotleft}$(c_1\mkern-0.5mu|\phi_1\rangle+c_2\mkern-0.5mu|\phi_1\rangle)\in(\mathcal{H}_{\neg\hat{Q}}\mkern-4mu\vee\mkern-2mu\mathcal{H}_{\hat{P}_n})^{\perp}\mkern1mu${\guillemotright} is false for any nonzero $c_1$ and $c_2$. This bring\smallskip

\begin{equation} 
   [\mkern-3.3mu[\neg{Q}\sqcup{P_n}]\mkern-3.3mu]
   =
   1
   -
   [\mkern-3.3mu[
      \text{\guillemotleft}
      (c_1\mkern-0.5mu|\phi_1\rangle+c_2\mkern-0.5mu|\phi_1\rangle)\in(\mathcal{H}_{\neg\hat{Q}}\mkern-4mu\vee\mkern-2mu\mathcal{H}_{\hat{P}_n})^{\perp}\mkern1mu
      \text{\guillemotright}
   ]\mkern-3.3mu]
   =
   1
   \;\;\;\;  ,
\end{equation}
\smallskip

\noindent therefore, the valuation (\ref{PART2}) returns $1=\min\{\mkern-3.3mu[P_1]\mkern-3.3mu],[\mkern-3.3mu[P_2]\mkern-3.3mu]\}$ meaning $[\mkern-3.3mu[{P_1}\sqcap{P_2}]\mkern-3.3mu]=1$.\\

\noindent As it turns out, the solution of the problem of an extra object discussed in \cite{Birkhoff, Piron} causes nondistributiveness of the lattice $L(\mathcal{H})$, namely,\smallskip

\begin{equation}  
   \mathcal{H}_{\neg\hat{Q}}
   \vee
   \big(
      \mathcal{H}_{\hat{P}_1}
      \wedge
      \mathcal{H}_{\hat{P}_2}
   \big)
   \neq
   \mkern-14mu
   \bigwedge_{n\in\{1,2\}}
      \mkern-16mu
      \left(
         \mathcal{H}_{\neg\hat{Q}}
         \vee
         \mathcal{H}_{\hat{P}_n}
      \right)
\;\;\;\;  ,
\end{equation}
\smallskip

\noindent which can be taken to be responsible for duality in the truth assignment for the propositional formula ${P_1}\sqcap{P_2}$.\\

\noindent One can assume from the above that even if both slits are open (and no which-way detector is placed behind them), the quantum particle will pass through just one slit, and so the exclusive disjunction ${P_1}\mkern4mu\underline{\sqcup}\mkern4mu{P_2}$ will always be true. Regarding the valuation $[\mkern-3.3mu[{P_1}\sqcap{P_2}]\mkern-3.3mu]=1$, one can blame it on the use of the distributivity law that does not hold generally in the Hilbert lattice.\\

\noindent From this assumption the ensuing hypothesis can be construed: A non-distributive logic (i.e., one where the statement $\neg{Q}\sqcup\mkern-0.5mu({P_1}\mkern2mu\sqcap\mkern2mu{P_2})\mkern-2mu\Leftrightarrow\mkern-2mu(\neg{Q}\mkern2mu\sqcup\mkern2mu{P_1})\sqcap(\neg{Q}\mkern2mu\sqcup\mkern2mu{P_2})$ does not need to be valid) underlies quantum phenomena and, for this reason, should be regarded as the correct logic for reasoning about the microscopic world \cite{Dickson}. In such a view, classical logic is merely a limiting case of a non-distributive logic \cite{Putnam68}. It is obvious that the minute the said hypothesis is accepted, the conundrum of wave–particle duality will cease to exist.\\

\noindent However, the aforesaid hypothesis gives rise to another problem, which can be posed as the following question: Given that the true logic is non-distributive, how may it be the case that the logical connectives $\sqcap$ and $\sqcup$ still distribute one over the other on some occasions?\\

\noindent The essence of this problem stated previously in the Section 2 (and discussed briefly in \cite{Bacciagaluppi}) is that it is impossible to determine – based solely on a propositional formula – in what circumstances the logical operators $\sqcap$ and $\sqcup$ will distribute. To do so, the meaning of propositions involved in the formula, specifically, the fact that the propositions have a classical content, must be taken into consideration. However, in propositional logic (including its modification such as a non-distributive logic), a propositional formula is concerned with the rules used for constructing an expression from atomic propositions and has nothing to do with an interpretation or meaning given to these propositions.\\

\section{The proposal of a partial bivaluation}  

\noindent The alternative proposal allowing to overcome the problem of an extra object is to assume that the function (\ref{MAP}) is a partial bijective function that associates elements of $\mathbb{B}_2$ with some elements of $\mathbb{B}_2\times\mathbb{B}_2$. Specifically, this function denotes the following:\smallskip

\begin{equation} \label{APBF} 
   \begingroup
   \begin{array}{r c c l}
      b:
      &
      \mathbb{B}_2\times\mathbb{B}_2
      &
      \cancel{\mkern-8mu\to\mkern-12mu}
      &
      \mathbb{B}_2
      \\[5pt]
      \hfill
      &
      (0,0)
      &
      \cancel{\mkern-8mu\mapsto\mkern-12mu}
      &
      \tau\in\mathbb{B}_2
   \end{array}
   \endgroup
   \;\;\;\;  ,
\end{equation}
\smallskip

\noindent where the first part can be read as: “$b$ is a partial function from $\mathbb{B}_2\mkern-2mu\times\mkern-2mu\mathbb{B}_2$ to $\mathbb{B}_2$”, while the second part is read: “$(0,0)$ does not map onto  anything in $\mathbb{B}_2$”. That is, $b$ is only defined on the pairs $(1,0)$ and $(0,1)$ whereas $b(0,0)$ stays undefined.\\

\noindent The application of this proposal (which can be called \emph{the proposal of a partial bivaluation}) to (\ref{PR1}) yields the following truth assignment:\smallskip

\begin{equation} \label{SV} 
   n
   \in
   \{1,2\}
   \mkern-3.3mu
   :
   \mkern6mu
   [\mkern-3.3mu[P_n]\mkern-3.3mu]
   =
   b
   \left(
      [\mkern-3.3mu[s_n]\mkern-3.3mu]
      ,
      [\mkern-3.3mu[s_{n-(-1)^n}]\mkern-3.3mu]
   \right)
   =
   \left\{
      \begingroup\SmallColSep
      \begin{array}{r l}
         1
         &
         \mkern3mu
         ,
         \mkern12mu
         [\mkern-3.3mu[s_n]\mkern-3.3mu]
         =
         1
         \\[3pt]
         0
         &
         \mkern3mu
         ,
         \mkern12mu
         [\mkern-3.3mu[s_{n-(-1)^n}]\mkern-3.3mu]
         =
         1
         \\[3pt]
         \textnormal{undefined}
         &
         \mkern3mu
         ,
         \mkern12mu
         [\mkern-3.3mu[s_n]\mkern-3.3mu]
         =
         [\mkern-3.3mu[s_{n-(-1)^n}]\mkern-3.3mu]
         =
         0
      \end{array}
      \endgroup   
   \right.
   \;\;\;\;  .
\end{equation}
\smallskip

\noindent As this assignment indicates, the lack of negative evidence (that is, the equivalence $[\mkern-3.3mu[s_{n-(-1)^n}]\mkern-3.3mu]=0$) does not guarantee that the proposition $P_n$ is true; even more so, when neither evidence is given (i.e., both $[\mkern-3.3mu[s_n]\mkern-3.3mu]$ and $[\mkern-3.3mu[s_{n-(-1)^n}]\mkern-3.3mu]$ are zero), this proposition has no truth value at all. Hence, the above assignment may be viewed as one done \emph{constructively}, i.e., using a semantics which only admits constructive proofs.\\

\noindent It is worthy of notice that a truth-value gap – i.e., lack of a truth value – does not stand for an intermediate truth value, akin to the value U (“undefined” or ``undetermined'' ) in a three-valued semantics.\\

\noindent To see this, let us interpret the truth-value gap as the indeterminate form $\frac{0}{0}$. In a loose manner of speaking, $\frac{0}{0}$ can take on the values $0$, $1$, or $\infty$. That is, the expression $\frac{0}{0}$ does not provide sufficient information to determine its value; in other words, it is undefined \cite{Thomas}.\\

\noindent Hence, it is impossible to say whether the form $\frac{0}{0}$ is greater than or equal to 0, or whether $\frac{0}{0}$ is less than or equal to 1. Likewise, it is impossible to say whether a truth-value gap is “truer” than or identical to F  or whether it is “falser” than or identical to T. This means that a truth-value gap cannot be an intermediate truth value. Correspondingly, a semantics allowing a truth-value gap has no interpretation as a three-valued semantics.\\

\noindent Pondering upon the relation between a truth-value gap and the epistemic predicates “verified” and “falsified”, it is possible to state the following: Seeing that “verified” and “falsified” are not identified with “true” and “false” (it is sufficient to say that a statement may be verified at one time and unverified at another, however, it may be true even at times when it was not verified; likewise, it may be false without being falsified \cite{Putnam75}), a statement may be verified or falsified at a certain time without being previously either true or false (as a result, from that time onward this statement would be thought to be either a true one or a false one). For example, providing reproducibility of the double-slit experiment, the proposition $P_n$, which has a truth-value gap when both slits are open, can be verified in one instance and falsified in another by positioning a which-way detector at the slits. With that, the occurrence of the proposition $P_n$ being verified/falsified would be subject to variations due to chance.\\

\noindent This gives a reason to assign a probability to one or the other of the epistemic predicates “verified” and “falsified”. Suppose that at some time, a proposition is verified by an experience (e.g., an observation or experiment). Understandably, such an outcome of the experience would be either certain or impossible if the proposition where to have a definite truth value, that is, either “true” or “false” respectively. On the other hand, where the proposition to have no truth value at all, the outcome “verified” would be neither certain nor impossible. Accordingly, the probability of being verified must be dispersion-free, i.e., either 1 or 0, for a proposition having a definite truth value and different from both 1 and 0 for a proposition having a truth-value gap. The above can be presented symbolically as follows\smallskip

\begin{equation}  
   \Pr[\text{\guillemotleft}P\text{ is verified}\text{\guillemotright}]
   =
   \left\{
      \begingroup\SmallColSep
      \begin{array}{r l}
         1
         &
         \mkern3mu
         ,
         \mkern12mu
         [\mkern-3.3mu[P]\mkern-3.3mu]
         =
         1
         \\[3pt]
         0
         &
         \mkern3mu
         ,
         \mkern12mu
         [\mkern-3.3mu[P]\mkern-3.3mu]
         =
         0
         \\[3pt]
         x\in(0,1)
         &
         \mkern2.5mu
         ,
         \mkern12mu
         [\mkern-3.3mu[P]\mkern-3.3mu]
         \mkern5mu
        \textnormal{is undefined}
      \end{array}
      \endgroup
   \right.
   \;\;\;\;  ,
\end{equation}
\smallskip

\noindent where $\Pr[\text{\guillemotleft}P\text{ is verified}\text{\guillemotright}]$ denotes the probability that a proposition $P$ will be verified by the experience sometime.\\

\noindent The application of the proposal (\ref{APBF}) to (\ref{PR2}) produces\smallskip

\begin{equation}  
   [\mkern-3.3mu[Q{\mkern2mu\lesseqgtr\mkern2mu}P]\mkern-3.3mu]
   =
   b
   \left(
      [\mkern-3.3mu[z]\mkern-3.3mu]
      ,
      [\mkern-3.3mu[w]\mkern-3.3mu]
   \right)
   =
   \left\{
      \begingroup\SmallColSep
      \begin{array}{r l}
         1
         &
         \mkern3mu
         ,
         \mkern12mu
         [\mkern-3.3mu[z]\mkern-3.3mu]
         =
         1
         \\[3pt]
         0
         &
         \mkern3mu
         ,
         \mkern12mu
         [\mkern-3.3mu[w]\mkern-3.3mu]
         =
         1
         \\[3pt]
         \textnormal{undefined}
         &
         \mkern3mu
         ,
         \mkern12mu
         [\mkern-3.3mu[z]\mkern-3.3mu]
         =
         [\mkern-3.3mu[w]\mkern-3.3mu]
         =
         0
      \end{array}
      \endgroup   
   \right.
   \;\;\;\;  .
\end{equation}
\smallskip

\noindent As it follows, if the mathematical statements $z$ and $w$ are false together, the proposition $Q{\mkern2mu\lesseqgtr\mkern2mu}P$ has no truth value. That is, in case the projection operators $\hat{Q}$ and $\hat{P}$ do not commute, the propositions $Q$ and $P$ are neither able nor unable to be true together. Consequently, one has no permission to say that the subspaces $\mathcal{H}_{\hat{Q}}$ and $\mathcal{H}_{\hat{P}}$ representing $Q$ and $P$ are either ordered or not ordered by the subset relation $\subseteq$. Otherwise stated, one may not say that $L(\mathcal{H})$ is a poset.\\

\noindent Nevertheless, consider a Boolean block, that is, a subset of $L(\mathcal{H})$, in which any two elements, say, the subspaces $\mathcal{H}_{\hat{Q}}$ and $\mathcal{H}_{\hat{P}}$, correspond to mutually compatible projection operators, $\hat{Q}$ and $\hat{P}$, respectively. Inside this block, either $[\mkern-3.3mu[z]\mkern-3.3mu]$ or $[\mkern-3.3mu[w]\mkern-3.3mu]$ is equal to 1; thus, the proposition $Q{\mkern2mu\lesseqgtr\mkern2mu}P$ is either true or false. From this it is possible to deduce that each Boolean block is a partially ordered subset of $L(\mathcal{H})$. So, the meet $\mathcal{H}_{\hat{Q}}\wedge\mathcal{H}_{\hat{P}}$ and the join $\mathcal{H}_{\hat{Q}}\vee\mathcal{H}_{\hat{P}}$ can be defined within every Boolean block. E.g.,\smallskip

\begin{equation} \label{SV1} 
   \hat{Q}\hat{P}
   =
   \hat{P}\hat{Q}
   :
   \mkern14mu
   \begingroup\SmallColSep
   \begin{array}{l}
      \mathcal{H}_{\hat{Q}}\wedge\mathcal{H}_{\hat{P}}
      =
      \mathcal{H}_{\hat{Q}}\cap\mathcal{H}_{\hat{P}}
      \\[6pt]
      \mathcal{H}_{\hat{Q}}\vee\mathcal{H}_{\hat{P}}
      =
      \left(
         \mathcal{H}_{\neg\hat{Q}}\cap\mathcal{H}_{\neg\hat{P}}
      \right)^{\perp}
   \end{array}
   \endgroup
   \;\;\;\;  .
\end{equation}
\smallskip

\noindent Outside Boolean blocks, i.e., in case $\hat{Q}\hat{P}\neq\hat{P}\hat{Q}$, the above operations are not defined. One can infer from this fact that propositional formulas constructed from propositions associated with noncompatible projection operators are not defined either.\enlargethispage{\baselineskip}\\

\noindent If the proposition $Q{\mkern2mu\lesseqgtr\mkern2mu}P$ has a truth value (i.e., if it happens inside a Boolean block), the mathematical statements\smallskip

\begin{equation}  
   \begingroup\SmallColSep
   \begin{array}{l l l}
      \text{\guillemotleft}
      \mkern1mu
      |\Psi\rangle
      \in
      \left(
         \mathcal{H}_{\hat{Q}}
         \wedge
         \mathcal{H}_{\hat{P}}
      \right)
      \mkern1mu
      \text{\guillemotright}
      \mkern-2mu
      &
      =
      &
      \mkern2mu
      s_{\wedge}
      \\[7pt]
      \text{\guillemotleft}
      \mkern1mu
      |\Psi\rangle
      \in
      \left(
         \mathcal{H}_{\hat{Q}}
         \wedge
         \mathcal{H}_{\hat{P}}
      \right)^{\perp}
      \mkern1mu
      \text{\guillemotright}
      \mkern-2mu
      &
      =
      &
      \mkern2mu
      s_{{\wedge}^{\perp}}
   \end{array}
   \endgroup   
   \;\;\;\;  ,
\end{equation}
\smallskip

\noindent which serve as positive and negative evidence, respectively, for the propositional formula ${Q}\sqcap{P}$, cannot be true together. Explicitly, for any nonzero vector $|\Psi\rangle$ of a Hilbert space $\mathcal{H}$, one finds that $(s_{\wedge}\sqcap s_{{\wedge}^{\perp}})\Leftrightarrow\bot$. Similarly, the statements\smallskip

\begin{equation}  
   \begingroup\SmallColSep
   \begin{array}{l l l}
      \text{\guillemotleft}
      \mkern1mu
      |\Psi\rangle
      \in
      \left(
         \mathcal{H}_{\hat{Q}}
         \vee
         \mathcal{H}_{\hat{P}}
      \right)
      \mkern1mu
      \text{\guillemotright}
      \mkern-2mu
      &
      =
      &
      \mkern2mu
      s_{\vee}
      \\[7pt]
      \text{\guillemotleft}
      \mkern1mu
      |\Psi\rangle
      \in
      \left(
         \mathcal{H}_{\hat{Q}}
         \vee
         \mathcal{H}_{\hat{P}}
      \right)^{\perp}
      \mkern1mu
      \text{\guillemotright}
      \mkern-2mu
      &
      =
      &
      \mkern2mu
      s_{{\vee}^{\perp}}
   \end{array}
   \endgroup   
   \;\;\;\;  ,
\end{equation}
\smallskip

\noindent which are positive and negative evidence, in that order, for the propositional formula ${Q}\sqcup{P}$, do not admit each other, i.e., $(s_{\vee} \sqcap s_{{\vee}^{\perp}})\Leftrightarrow\bot$, if $|\Psi\rangle\mkern-2mu\in\mkern-2mu\mathcal{H}$ and $|\Psi\rangle\neq0$. Hence, in line with the proposal (\ref{APBF}), ${Q}\sqcap{P}$ and ${Q}\sqcup{P}$ may be valuated using only positive evidence, namely,\smallskip

\begin{equation} \label{SV2} 
   \hat{Q}\hat{P}
   =
   \hat{P}\hat{Q}
   :
   \mkern10mu
   \begingroup\SmallColSep
   \begin{array}{r}
      [\mkern-3.3mu[{Q}\sqcap{P}]\mkern-3.3mu]
      =
      [\mkern-3.3mu[s_{\wedge}]\mkern-3.3mu]
      \\[7pt]
      [\mkern-3.3mu[{Q}\sqcup{P}]\mkern-3.3mu]
      =
      [\mkern-3.3mu[s_{\vee}]\mkern-3.3mu]
   \end{array}
   \endgroup
   \;\;\;\;  .
\end{equation}
\smallskip

\noindent Thus, allowing the statement $Q{\mkern2mu\lesseqgtr\mkern2mu}P$ to be partially valuated is equivalent to \emph{giving up the lattice condition}: As a result of doing this, the logical connectives $Q{\mkern2mu\sqcap\mkern2mu}P$ and $Q{\mkern2mu\sqcup\mkern2mu}P$ turn out to exist only for countable sets of propositions $Q$ and $P$ that are represented by pairwise orthogonal or colinear subspaces of the Hilbert space of the system (by contrast, in the works \cite{Chiara} and \cite{Griffiths}, the abandonment of the lattice condition is suggested as an original assumption).\\

\noindent Recall that as per the formula (\ref{P}), the projection operators $\hat{P}_1$ and $\hat{P}_2$ representing the atomic propositions $P_1$ and $P_2$ in the double-slit experiment are compatible and orthogonal. Moreover, according to the completeness relation $\hat{P}_1+\hat{P}_2=\hat{1}$, the sum of their corresponding subspaces $\mathcal{H}_{\hat{P}_1}$ and $\mathcal{H}_{\hat{P}_2}$ is the identical subspace $\mathcal{H}$; in symbols, $\mathcal{H}_{\hat{P}_1}+\mathcal{H}_{\hat{P}_2}=\mathcal{H}$. So, the meet and join of the subspaces $\mathcal{H}_{\hat{P}_1}$ and $\mathcal{H}_{\hat{P}_2}$ can be defined by\smallskip

\begin{equation}  
   \hat{P}_1\hat{P}_2
   =
   \hat{P}_2\hat{P}_1
   =
   0
   :
   \mkern14mu
   \begingroup\SmallColSep
   \begin{array}{l}
      \mathcal{H}_{\hat{P}_1}\wedge\mathcal{H}_{\hat{P}_2}
      =
      \{0\}
      \\[6pt]
      \mathcal{H}_{\hat{P}_1}\vee\mathcal{H}_{\hat{P}_2}
      =
      \left(
         \mathcal{H}_{\hat{P}_2}\wedge\mathcal{H}_{\hat{P}_1}
      \right)^{\perp}
      \mkern-8mu
      =
      \mathcal{H}
   \end{array}
   \endgroup
   \;\;\;\;  .
\end{equation}
\smallskip

\noindent From the above one finds that the mathematical statements {\guillemotleft}$\mkern1mu|\Psi\rangle\mkern-4mu\in\mkern-4mu(\mathcal{H}_{\hat{P}_1}\mkern-4mu\wedge\mkern-0.5mu\mathcal{H}_{\hat{P}_2})\mkern1mu${\guillemotright} and {\guillemotleft}$\mkern1mu|\Psi\rangle\mkern-4mu\in\mkern-4mu(\mathcal{H}_{\hat{P}_1}\mkern-4mu\vee\mkern-0.5mu\mathcal{H}_{\hat{P}_2})\mkern1mu${\guillemotright} are false and true, respectively, for any non-zero vector $|\Psi\rangle$ in the Hilbert space $\mathcal{H}$. Hence, at one with (\ref{SV2}), the following must hold\smallskip

\begin{equation}  
   \forall
   |\Psi\rangle\mkern-2mu
   \in
   \mathcal{H}
   \backslash
   \{0\}
   :
   \mkern10mu
   \begingroup\SmallColSep
   \begin{array}{r}
      [\mkern-3.3mu[{P_1}\sqcap{P_2}]\mkern-3.3mu]
      =
      [\mkern-3.3mu[
         \text{\guillemotleft}\mkern1mu|\Psi\rangle\mkern-4mu\in\mkern-4mu(\mathcal{H}_{\hat{P}_1}\mkern-4mu\wedge\mkern-0.5mu\mathcal{H}_{\hat{P}_2})\mkern1mu\text{\guillemotright}
      ]\mkern-3.3mu]
      =
      0
      \\[7pt]
      [\mkern-3.3mu[{P_1}\sqcup{P_2}]\mkern-3.3mu]
      =
      [\mkern-3.3mu[
         \text{\guillemotleft}\mkern1mu|\Psi\rangle\mkern-4mu\in\mkern-4mu(\mathcal{H}_{\hat{P}_1}\mkern-4mu\vee\mkern-0.5mu\mathcal{H}_{\hat{P}_2})\mkern1mu\text{\guillemotright}
      ]\mkern-3.3mu]
      =
      1
   \end{array}
   \endgroup
   \;\;\;\;  .
\end{equation}
\smallskip

\noindent The projection operator $\hat{0}$, which is in the one-to-one correspondence with the subspace $\mathcal{H}_{\hat{P}_1}\mkern-4mu\wedge\mkern-0.5mu\mathcal{H}_{\hat{P}_2}$ representing ${P}_1\sqcap{P}_2$, is compatible with any projection operator; thus, the propositional formula $\neg{Q}\sqcup({P}_1\sqcap{P}_2)$ is defined and has the same truth value as the negation $\neg{Q}$ does. To be sure,\smallskip

\begin{equation}  
   (\hat{1}-\hat{Q})\mkern1.5mu\hat{0}
   =
   \hat{0}\mkern1.5mu(\hat{1}-\hat{Q})
   :
   \mkern10mu
   [\mkern-3.3mu[\neg{Q}\sqcup({P}_1\sqcap{P}_2)]\mkern-3.3mu]
   =
   \left\{
      \begingroup\SmallColSep
      \begin{array}{r l}
         1
         &
         \mkern3mu
         ,
         \mkern12mu
         [\mkern-3.3mu[
           \text{\guillemotleft}\mkern1mu|\Psi\rangle\mkern-4mu\in\mkern-4mu\mathcal{H}_{\neg\hat{Q}}\mkern1mu\text{\guillemotright}
         ]\mkern-3.3mu]
         =
         1
         \\[7pt]
         0
         &
         \mkern3mu
         ,
         \mkern12mu
         [\mkern-3.3mu[
            \text{\guillemotleft}\mkern1mu|\Psi\rangle\mkern-4mu\in\mkern-4mu\mathcal{H}_{\hat{Q}}\mkern1mu\text{\guillemotright}
         ]\mkern-3.3mu]
         =
         1
      \end{array}
      \endgroup
   \right.
   \;\;\;\;  .
\end{equation}
\smallskip

\noindent On the other hand, the projection operator $\hat{Q}$ corresponding to the subspace $\mathcal{H}_{\hat{Q}}$ (which contains the vector $|\Psi\rangle=c_1|\phi_1\rangle+c_2|\phi_2\rangle$ where $|\phi_1\rangle\in\mathcal{H}_{\hat{P}_1}$ and $|\phi_2\rangle\in\mathcal{H}_{\hat{P}_2}$) is compatible with neither $\hat{P}_1$ nor $\hat{P}_2$. The same holds for $\neg{\hat{Q}}$ and $\hat{P}_n$: $(\hat{1}-\hat{Q})\hat{P}_n\neq\hat{P}_n(\hat{1}-\hat{Q})$. Given that the subspaces $\mathcal{H}_{\neg\hat{Q}}$ and $\mathcal{H}_{\hat{P}_n}$ cannot be in one Boolean block, the operation $\mathcal{H}_{\neg\hat{Q}}\wedge\mathcal{H}_{\hat{P}_n}$ is not defined and therefrom the propositional formula $(\neg{Q}\sqcup{P_1})\sqcap(\neg{Q}\sqcup{P_2})$ is not defined either.\\

\noindent Note the difference between the false statement $\neg{Q}\sqcup\mkern-0.5mu({P_1}\mkern2mu\sqcap\mkern2mu{P_2})\Leftrightarrow(\neg{Q}\mkern2mu\sqcup\mkern2mu{P_1})\sqcap(\neg{Q}\mkern2mu\sqcup\mkern2mu{P_2})$ and the meaningless statement $\neg{Q}\sqcup\mkern-0.5mu({P_1}\mkern2mu\sqcap\mkern2mu{P_2})\Leftrightarrow{n.d.f.}$, where $n.d.f.$ stays for ``not defined formula''. The negation of the former yields ${Q}\sqcap\mkern-0.5mu({P_2}\mkern2mu\sqcup\mkern2mu{P_1})\nLeftrightarrow({Q}\mkern2mu\sqcap\mkern2mu{P_2})\sqcup({Q}\mkern2mu\sqcap\mkern2mu{P_1})$, that is, $Q\nLeftrightarrow{\bot}$, which is true, whereas the negation of the latter returns $Q\nLeftrightarrow{n.d.f.}$, which is meaningless once again. The said difference can be interpreted as that unlike the proposal suggested by Birkhoff and von Neumann, the one that assumes a partial bivaluation does not entail the failure of distributivity.\\

\noindent Let either of slits be open (or let a which-way detector register the particle at one or the other slit). Then, in accordance with (\ref{SV}), one of the propositions $P_n$ is true while the other is false, so that the statement ${P_1}\mkern2mu\underline{\sqcup}\mkern2mu{P_2}$ is a tautology. This implies that in the given case the logical connectives $\sqcap$ and $\sqcup$ are truth-functional.\\

\noindent Now, suppose that both slits are open, and no which-way detector is placed at the double-slit plate. In that case, the operators of logical conjunction and disjunction cannot be said to be truth-functional. In fact, even though both $P_1$ and $P_2$ happen to have a truth-value gap when both slits are open, ${P_1}\mkern2mu\sqcup\mkern2mu{P_2}$ and ${P_1}\mkern2mu\sqcap\mkern2mu{P_2}$ continue to be true and false, respectively, and the statement ${P_1}\mkern2mu\underline{\sqcup}\mkern2mu{P_2}$ remains necessary true, i.e., it stays such that no instance exists in which this statement could fail to be true:\smallskip

\begin{equation} \label{PROOF} 
   \forall
   |\Psi\rangle\mkern-2mu
   \in
   \mathcal{H}
   \backslash
   \{0\}
   :
   \mkern10mu
   [\mkern-3.3mu[
      {P_1}\mkern2mu\underline{\sqcup}\mkern2mu{P_2}
   ]\mkern-3.3mu]
   =
   \min\mkern-2mu\Big\{
       [\mkern-3.3mu[
          \text{\guillemotleft}\mkern0.5mu|\Psi\rangle\mkern-4mu\in\mkern-4mu\mathcal{H}\mkern1mu\text{\guillemotright}
       ]\mkern-3.3mu]
      \mkern1mu
      ,
      \mkern1.5mu
      1
      -
       [\mkern-3.3mu[
          \text{\guillemotleft}\mkern0.5mu|\Psi\rangle\mkern-4mu\in\mkern-4mu\{0\}\mkern1mu\text{\guillemotright}
       ]\mkern-3.3mu]
   \big\}
   =
   1
   \;\;\;\;  .
\end{equation}
\smallskip

\noindent Hence, the proposal based on the assumption of a partial bivaluation does not bring about the wave-particle duality. As stated by this proposal, the quantum particle can always be described as a particle-like thing, regardless of determination of a measuring device in the double-slit experiment.\\

\section{Concluding remarks}  

\noindent Birkhoff and von Neumann’s proposal assumes that without negative evidence – i.e., when the mathematical statement $s_{n-(-1)^n}$ is false – the proposition $P_n$ should be accepted as a true one. In other words, according to this proposal, the absence of a demonstration that a proposition is false guarantees that the proposition is true. Clearly, such an assumption would be correct if every proposition were to have a definite truth value regardless of proof.\\

\noindent On the other hand, in a semantics validating classical logic, every proposition is conceived as possessing a determinate truth value independently of whether we know it or have at our disposal the means to prove it \cite{Dummett}. One can conclude from this that a semantics validating Birkhoff and von Neumann’s proposal is like a semantics that bears out classical logic.\\

\noindent But here lies the irony: Birkhoff and von Neumann’s proposal, which was intended to be a replacement for classical logic in the domain of quantum mechanics, has at its core a semantics of classical logic that underlies the principles of classical mechanics. This may explain why Birkhoff and von Neumann’s proposal for quantum logic has not made a great deal of progress in a solution of the quantum conceptual difficulties.\\

\noindent Contrastively, a semantics validating the proposal of a partial bivaluation differs from any semantics that classical logic might have (at least in some crucial respect). Namely, in the said semantics (called \emph{supervaluationism}), a propositional formula can possess a definite truth value even if its constituent propositions do not \cite{Williamson, Keefe, Varzi}. Hence, supervaluation semantics may offer a different approach to the quantum conceptual problems.\\

\noindent For example, according to the textbook interpretation of quantum mechanics \cite{Faye}, our choice of what to observe in the double-slit experiment determines the properties of a quantum particle therein. Accordingly, the quantum particle stops behaving like a wave and becomes a particle-like entity when an observation of a particle's path takes place.\\

\noindent But according to the proposal of a partial bivaluation, the quantum particle’s properties are not contingent upon observation. To be exact, in the double-slit experiment, the quantum particle always behaves like a particle-like entity – i.e., one that goes along one path or the other but not both – irrespective of the observation. Together with all that, a statement affirming that the quantum particle follows a certain path has no truth value at all. One can only consider the probability that this statement will be verified if the actual particle's path is observed. Because of that, behavior of a quantum particle emerging from the double-slit plate is impossible to convey in the classical concepts – i.e., ones that are based on classical logic.\\

\noindent The similar conclusion is reached in \cite{Trassinelli}: The statements {\guillemotleft}The particle passed through the slit 1{\guillemotright} and {\guillemotleft}The particle passed through the slit 2{\guillemotright} are completely indeterminate if we do not measure where the particle passed through. However, the premise, from which this conclusion has been inferred, is totally different from that stated in the present paper. Here, the truth or the falsity of any statement is assumed to be not primitive but derivative of evidence; as a result, a statement has no truth value if no evidence exists attesting that the statement is true or false. By way of contrast, in \cite{Trassinelli} the indeterminate status of a statement happens on account of the special definition of the conditional probability for measurement outcomes (this probability is defined in a such way that the distributive law need not be valid).\\

\noindent Furthermore, the application of the standard interpretation of quantum mechanics to any sentient creature leads to a paradox known as Schrödinger’s cat. The paradox involves a cat which – in agreement with the aforesaid interpretation – may be dead and alive at the same time. It could remain in such an inconceivable and absurd state for an arbitrarily long period of time, until the observer opens an opaque box enclosing the cat, at which point the animal is either dead or alive.\\

\noindent In contrast, according to the proposal of a partial bivaluation, the cat (together with a decaying radioactive atom on which its fate depends) is in either one state or another but never both, regardless of a “conscious” or “unconscious” observation That is, for the cat, the premise of “macroscopic realism” \cite{Leggett} (declaring that a macroscopic system is in one or other of two macroscopically distinct states available to it but not in both) is always true. Therewithal, the proposition asserting that the cat is in a certain state (e.g., dead) prior to the observation has absolutely no truth value (because neither evidence exists for this proposition before the observation). Still, this proposition may be verified/falsified (with some probability) by opening the box.\\

\noindent Giving the great conceptual value of Wigner’s thought experiment \cite{Wigner}, it is worth discussing – even in brief – how the proposal of a partial bivaluation explains Wigner's conundrum of the friend.\\

\noindent Let us consider a slightly modified version of the standard account of the Wigner-friend thought experiment. This version posits an inside observer, a friend of Wigner, who has been asked to perform the double-slit experiment in a completely closed laboratory (so that an outside observer, Wigner, cannot be aware of anything happening in it until its door is open). Suppose that before the moment when the laboratory is closed, both Wigner and his friend agree that the statement ${P_1}\mkern2mu\underline{\sqcup}\mkern2mu{P_2}$ is a contradiction. Also suppose that the afterwards the inside observer introduces a which-way detector at the double-slit plate. Thenceforth, for this observer the statement ${P_1}\mkern2mu\underline{\sqcup}\mkern2mu{P_2}$ is a tautology. However, for the period that the door of the laboratory keeps on being closed, the same statement ${P_1}\mkern2mu\underline{\sqcup}\mkern2mu{P_2}$ will continue to be a contradiction for the outside observer. Denoting truth values, which Wigner’s friend and Wigner assign to an arbitrary statement, as $[\mkern-3.3mu[\cdot]\mkern-3.3mu]_{\text{F}}$  and $[\mkern-3.3mu[\cdot]\mkern-3.3mu]_{\text{W}}$, respectively, one finds\smallskip

\begin{equation} \label{INEQ} 
   [\mkern-3.3mu[{P_1}\mkern2mu\underline{\sqcup}\mkern2mu{P_2}]\mkern-3.3mu]_{\text{F}}
   \neq
   [\mkern-3.3mu[{P_1}\mkern2mu\underline{\sqcup}\mkern2mu{P_2}]\mkern-3.3mu]_{\text{W}}
   \;\;\;\;  .
\end{equation}
\smallskip

\noindent In this way, the application of \emph{the instrumentalist description of quantum mechanics} (i.e., the description which merely relates the mathematical formalism of quantum theory to data and prediction) \emph{alongside classical logic} appears to show that the statement ${P_1}\mkern2mu\underline{\sqcup}\mkern2mu{P_2}$ is paradoxical: At one and the same time, it is true for one observer and false for the other.\\

\noindent To make this paradox even more dramatic, in Frauchiger and Renner's thought experiment \cite{Frauchiger}, Wigner uses two different methods to assign a truth value to the statement ${P_1}\mkern2mu\underline{\sqcup}\mkern2mu{P_2}$. When he uses the standard quantum formalism (i.e., the instrumentalist description of quantum mechanics in conjunction with classical logic), he gets $[\mkern-3.3mu[{P_1}\mkern2mu\underline{\sqcup}\mkern2mu{P_2}]\mkern-3.3mu]_{\text{W}}=0$ (as stated above). But when he reasons about what truth value his friend might have assigned to this statement, Wigner may decide $[\mkern-3.3mu[{P_1}\mkern2mu\underline{\sqcup}\mkern2mu{P_2}]\mkern-3.3mu]_{\text{W}}=1$ or $[\mkern-3.3mu[{P_1}\mkern2mu\underline{\sqcup}\mkern2mu{P_2}]\mkern-3.3mu]_{\text{W}}=0$. Whatever the case may be, when the second method is used, $[\mkern-3.3mu[{P_1}\mkern2mu\underline{\sqcup}\mkern2mu{P_2}]\mkern-3.3mu]_{\text{W}}$ cannot be a certain value, in direct contradiction to the unconditional status of the statement ${P_1}\mkern2mu\underline{\sqcup}\mkern2mu{P_2}$. As Frauchiger and Renner argue, such a contradiction indicates that quantum theory cannot be extrapolated to complex systems.\\

\noindent Wigner's puzzle takes us back to the central question of the present paper: Are the truth values “true” and “false” absolute, or relative to observers?\\

\noindent Based on what is proposed in \cite{Baumann}, the inequality (\ref{INEQ}) could be dissolved if the inside observer were regarded as \emph{a rational agent} (i.e., an entity like a team of scientists sharing notebooks, calculations, observations, etc., who may freely take actions on parts of the world external to themselves \cite{Stacey}). In that case, Wigner’s friend might believe that assigning truth values to the statement ${P_1}\mkern2mu\underline{\sqcup}\mkern2mu{P_2}$ conditioned solely to events happening inside the laboratory would not be rational, and consequentially the valuation $[\mkern-3.3mu[{P_1}\mkern2mu\underline{\sqcup}\mkern2mu{P_2}]\mkern-3.3mu]_{\text{F}}=1$ would not be correct.\\

\noindent But according to both quantum Bayesianism (abbreviated QBism) \cite{Fuchs} and the relational quantum mechanics (abbreviated RQM) \cite{Rovelli, Biagio}, the paradoxicality of the truth assignment for ${P_1}\mkern2mu\underline{\sqcup}\mkern2mu{P_2}$ disappears when “true” and “false” are identified with the “experiences” of different observers. In both QBism and RQM, “true” and “false” are not objective, rather “facts relative to the observers” – in other words, they are observers’ personal judgements \cite{DeBrota}. In a slogan: `One’s “true” might be someone else's “false”'.\\

\noindent This by no means indicates that QBism and RQM are self-contradictory: The moment the door of the laboratory is open for a second time, Wigner and his friend (along with the whole laboratory) will combine into one system, and so Wigner‘s “false” will be converted into friend’s “true”. Consequently, Wigner and his friend will never be able to prove a disagreement between their truth assignments.\\

\noindent Though the case be such, not only does the dependency of the truth values upon an agent amount to giving up the absolute nature of facts, but most importantly it has the value of forsaking the objectivity of mathematics. That is, if “true” and “false” are subject to agent’s judgements, then the truth of every mathematical statement must be an agent’s belief, “supremely strong, but nonetheless a belief”.\\

\noindent The question regarding the objectivity of mathematics, such as whether mathematical truth is objective or subjective, is perhaps one of the oldest and hardest questions in Western philosophy.\\

\noindent On the one hand, seeing as mathematics is a free activity of the mind, one may consider mathematical truth subjective. In agreement with Putnam \cite{Putnam75}, if the only method allowed in mathematics had consisted in deriving conclusions from axioms, which have been fixed permanently and for all possible agents (using mathematics), then the truth of any mathematical statement would have been objective, i.e., independent of agents’ personal judgements. But it did not. To prove a mathematical statement an agent may use empirical and probabilistic arguments seemed plausible for this agent but not acceptable for the other. As a result, the mathematical statement in question would be true for the former and false for the latter. For example, until a well-defined meaning was given to the Dirac delta function, $\delta\mkern-1mu(\mkern-1mu{x}\mkern-2mu)$, the computations made using this function appeared to most mathematicians as nonsense \cite{Bracewell}. Consequently, even though a mathematical statement involving this function such as \text{\guillemotleft}$(x\mkern1.5mu\delta\mkern-1mu(\mkern-1mu{x}\mkern-2mu)=0)\sqcap(\int_{-\infty}^{+\infty}\delta\mkern-1mu(\mkern-1mu{x}\mkern-2mu)\mkern1.5mu\mathrm{d}x=1)$\text{\guillemotright} might be considered true by some mathematicians, it would be regarded as false by the rest.\\

\noindent Despite this, if a mathematical statement is true because it has a constructive proof, such a statement will be true absolutely. Indeed, in that case there is a constructive witness to the truthfulness of the statement, that is, an actual example proving it true. And because this example exists in fact and not merely potential or possible, it is the fact for all agents. Hence, the said mathematical statement will be true for every agent, that is, true absolutely.\\

\noindent From the above reasoning it follows that by accepting the relativity of the truth values QBism and RQM merely reject constructivist philosophy.\\

\noindent That is quite different from the proposal of a partial bivaluation. To be sure, let us consider the mathematical statement {\guillemotleft}$|\Psi\rangle\mkern-2mu\in\mathcal{H}\backslash\{0\}${\guillemotright}. Any physically meaningful state, i.e., every state describable by non-zero vector $|\Psi\rangle$ of a finite-dimensional Hilbert space $\mathcal{H}$, is computable. Therefore, as far as a physical meaning is concerned, there exists computational evidence witnessing the truth of {\guillemotleft}$|\Psi\rangle\mkern-2mu\in\mathcal{H}\backslash\{0\}${\guillemotright}. What is more, because computability of $|\Psi\rangle$ implies a constructive mechanism generating $|\Psi\rangle$, one may say that {\guillemotleft}$|\Psi\rangle\mkern-2mu\in\mathcal{H}\backslash\{0\}${\guillemotright} is constructively true.\\

\noindent This entails the absolute truth of ${P_1}\mkern2mu\underline{\sqcup}\mkern2mu{P_2}$ whose evidence, in accordance with (\ref{PROOF}), is provided by the truthfulness of {\guillemotleft}$|\Psi\rangle\mkern-2mu\in\mathcal{H}\backslash\{0\}${\guillemotright}. Consequently, the truth of the unconditional statement {\guillemotleft}In the double-slit experiment, the quantum particle passes through one or the other slit, but not both{\guillemotright} must be shared alike by Wigner’s friend and Wigner:\smallskip

\begin{equation} \label{OIT} 
   [\mkern-3.3mu[{P_1}\mkern2mu\underline{\sqcup}\mkern2mu{P_2}]\mkern-3.3mu]_{\text{F}}
   =
   [\mkern-3.3mu[{P_1}\mkern2mu\underline{\sqcup}\mkern2mu{P_2}]\mkern-3.3mu]_{\text{W}}
   =
   1
   \;\;\;\;  ,
\end{equation}
\smallskip

\noindent i.e., it must be an observer-independent fact.\\

\noindent Using the terminology of Brukner’s no-go theorem \cite{Brukner}, one can declare that the above equality implies the compatibility of “universal validity of quantum theory” (that is, the applicability of the mathematical formalism of the theory to any physical system without restriction), “locality” (i.e., the independence of one observer’s measurement settings and other observer’s outcomes) and “freedom of choice” (meaning that Wigner’s friend can freely decide to introduce or not to introduce a which-way detector) with the assumption of “observer-independent facts” (under this assumption, $[\mkern-3.3mu[{P_1}\mkern2mu\underline{\sqcup}\mkern2mu{P_2}]\mkern-3.3mu]_{\text{F}}$ and $[\mkern-3.3mu[{P_1}\mkern2mu\underline{\sqcup}\mkern2mu{P_2}]\mkern-3.3mu]_{\text{W}}$ may be defined together).\\

\noindent Now, let us consider the valuation of the atomic propositions $P_n$. For Wigner, these propositions have no truth values at all, accordingly, the set of $[\mkern-3.3mu[{P_n}]\mkern-3.3mu]_{\text{W}}$ is $\varnothing$, the set with no elements. At the same time, $P_n$ are verified/falsified by his friend and, as a result, the set of $[\mkern-3.3mu[{P_n}]\mkern-3.3mu]_{\text{F}}$ is not $\varnothing$. In symbols:\smallskip

\begin{equation} \label{OIP} 
   \bigg\{
      n
      \mkern-3mu
      \in
      \mkern-3mu
      \{1,2\}
      \textnormal{:}
      \mkern10mu
      [\mkern-3.3mu[{P_n}]\mkern-3.3mu]_{\text{F}}
   \bigg\}
   \mkern5mu
   \neq
   \mkern5mu
   \bigg\{
      n
      \mkern-3mu
      \in
      \mkern-3mu
      \{1,2\}
      \textnormal{:}
      \mkern10mu
      [\mkern-3.3mu[{P_n}]\mkern-3.3mu]_{\text{W}}
   \bigg\}
   \;\;\;\;  .
\end{equation}
\smallskip

\noindent This means that one cannot define $[\mkern-3.3mu[{P_n}]\mkern-3.3mu]_{\text{F}}$ and $[\mkern-3.3mu[{P_n}]\mkern-3.3mu]_{\text{W}}$ together. Moreover, one cannot define a joint probability $p_{\text{joint}}$ such that\smallskip

\begin{equation}  
   p_{\text{joint}}
   =
   \mathrm{Pr}
   \bigg[
      \text{\guillemotleft}
      [\mkern-3.3mu[{P_n}]\mkern-3.3mu]_{\text{F}}
      \in
      \{0,1\}
      \mkern6mu      
      \textnormal{\bf{and}}
      \mkern6mu      
      [\mkern-3.3mu[{P_n}]\mkern-3.3mu]_{\text{W}}
      \in
      \{0,1\}
   \text{\guillemotright}
   \bigg]
   \;\;\;\;  ,
\end{equation}
\smallskip

\noindent and neither can one define a truth-valued function corresponding to a binary connective (say, conjunction) that joins the proposition $P_n$ evaluated by Wigner’s friend with the proposition $P_n$ evaluated by Wigner. In this way, the truth or falsity of the statement {\guillemotleft}The quantum particle passes through the slit $n${\guillemotright} is a fact only for a particular observer.\\

\noindent Looking at (\ref{OIT}) and (\ref{OIP}), one can infer that quantum theory treats facts in a dual manner: On the one hand, in this theory statements known to be true or false from an experience of an observer are facts relative to this observer. But on the other hand, quantum theory also grants entrance to observer-independent facts, i.e., statements which are true or false for all observers without regard for their experiences or beliefs. Such a mode of action can be taken as an indication or sign of \emph{the dual – i.e., objective-subjective – nature of facts in quantum theory}.\\

\noindent In the view of that nature, Brukner’s no-go theorem, which asserts that in quantum theory one can only define facts relative to an observation and an observer, might be applicable only in part. The same holds for a new strong no-go theorem by Bong et al \cite{Bong} built on assumptions strictly weaker than those of Brukner’s no-go theorem.\\

\noindent As to the recent attempt by Proietti et al \cite{Proietti} to perform an extended Wigner’s friend thought experiment (with four observers), the following may be remarked in passing. Performing a thought experiment is not a question of technical capabilities or ingenuity. Even if it could be possible to perform a thought experiment, there need not be an intention (a motive or a purpose) to perform it. This is so because in thought experiments one gains new information not by observing or measuring events or experiences but by reorganizing or rearranging \emph{already known experimental data} in a new way with the aim of drawing new inference from them \cite{Brown}. So, regarding the Wigner’s friend thought experiment (the original version or an extended modification of it), one already has all the necessary empirical data (provided by almost 100 years of quantum mechanics studies) allowing one to formulate statements about physical systems' states after the measurement. Furthermore, the purpose of the said experiment is to challenge (or even refute) the Copenhagen interpretation of quantum mechanics (using the device of the imagination known as reductio ad absurdum). Given that the Copenhagen interpretation is \emph{unfalsifiable} (no evidence that comes to life can contradict it), any attempt to perform the Wigner’s friend experiment in real life is merely fruitless.\\

\section*{Acknowledgement}  

\noindent The author wishes to thank the anonymous referee for the inspiring remarks and constructive comments which helped him enrich and deepen this paper.\\

\bibliographystyle{References}

\begin{thebibliography}{10}
\expandafter\ifx\csname urlstyle\endcsname\relax
  \providecommand{\doi}[1]{doi:\discretionary{}{}{}#1}\else
  \providecommand{\doi}{doi:\discretionary{}{}{}\begingroup
  \urlstyle{rm}\Url}\fi

\bibitem{Greiner}
Walter Greiner and David~A. Bromley.
\newblock \emph{Quantum {M}echanics: {A}n {I}ntroduction ({T}heoretical
  physics)}.
\newblock Springer, 2012.

\bibitem{Bohr}
Niels Bohr.
\newblock The quantum postulate and the recent development of atomic theory.
\newblock \emph{Nature {S}upplement}, 121:580--590, 1928.

\bibitem{Scully}
Marian~O. Scully, Berthold-Georg Englert, and Herbert Walther.
\newblock Quantum optical tests of complementarity.
\newblock \emph{Nature}, 351:111--116, 1991.

\bibitem{Menzel}
Ralf Menzel, Dirk Puhlmann, Axel Heuer, and Wolfgang~P. Schleich.
\newblock Wave-particle dualism and complementarity unraveled by a different
  mode.
\newblock \emph{Proceedings of National Academy of Sciences of USA},
  109(24):9314--9319, 2012.

\bibitem{Feynman}
Richard~P. Feynman, Robert~B. Leighton, and Matthew Sands.
\newblock \emph{The {F}eynman {L}ectures on {P}hysics, volume 1}.
\newblock Addison-Wesley, 1963.

\bibitem{Birkhoff}
Garrett Birkhoff and John von Neumann.
\newblock The logic of quantum mechanics.
\newblock \emph{Annals of Mathematics}, 37:823--843, 1936.

\bibitem{Baltag}
Alexandru Baltag and Sonja Smets.
\newblock Quantum logic as a dynamic logic.
\newblock \emph{Synthese}, 179:285--306, 2011.

\bibitem{Mateus}
Paulo Mateus and Amilcar Sernadas.
\newblock Weakly complete axiomatization of exogenous quantum propositional
  logic.
\newblock \emph{Information and Computation}, 204:771--794, 2006.

\bibitem{Abramsky}
Samson Abramsky and Ross Duncan.
\newblock A categorical quantum logic.
\newblock \emph{Math. Struct. Comp. Science}, 16:469--489, 2006.

\bibitem{Pavicic}
Mladen Pavi{$\check{\mathrm{c}}$}i{$\acute{\mathrm{c}}$}.
\newblock Bibliography on quantum logics and related structures.
\newblock \emph{Int. J. Theor. Phys}, 31:373--461, 1992.

\bibitem{Kochen}
Simon Kochen and Ernst Specker.
\newblock Logical structures arising in quantum mechanics.
\newblock In Gerhard J{$\ddot{\mathrm{a}}$}ger, Hans
  L{$\ddot{\mathrm{a}}$}uchli, Bruno Scarpellini, and Volker Strassen, editors,
  \emph{Ernst Specker Selecta}, pages 210--221. Birkh{$\ddot{\mathrm{a}}$}user
  Verlag Basel, 1990.

\bibitem{Specker}
Simon Kochen and Ernst Specker.
\newblock The calculus of partial propositional functions.
\newblock In Clitford Hooker, editor, \emph{The {L}ogico-algebraic approach to
  quantum mechanics: {H}istorical {E}volution}, volume~1, pages 277--292. D.
  Reidel Publishing Company, P.O. Box 17, Dordrecht, Holland, 1974.

\bibitem{Kochen15}
Simon Kochen.
\newblock A {R}econstruction of {Q}uantum {M}echanics.
\newblock \emph{Found Phys}, 45:557--590, 2015.

\bibitem{Church}
Alonzo Church.
\newblock \emph{Introduction to {M}athematical {L}ogic}.
\newblock Princeton University Press, Princeton, NJ, 1956.

\bibitem{Mendelson}
Elliott Mendelson.
\newblock \emph{Introduction to {M}athematical {L}ogic}.
\newblock Springer, 1997.

\bibitem{Klement}
Kevin~C. Klement.
\newblock Propositional logic.
\newblock In \emph{The Internet Encyclopedia of Philosophy}.
  {https://www.iep.utm.edu/prop-log/}, 2020.

\bibitem{Mirsky}
Leonid Mirsky.
\newblock \emph{An {I}ntroduction to {L}inear {A}lgebra. {D}over {B}ooks on
  {M}athematics}.
\newblock Dover Publications, 2011.

\bibitem{Sawant}
Rahul Sawant, Joseph Samuel, Aninda Sinha, Supurna Sinha, and Urbasi Sinha.
\newblock Non-classical paths in interference experiments.
\newblock \emph{Physical Review Letters}, 113(12):120406, 2014.

\bibitem{Neumann}
John von Neumann.
\newblock \emph{Mathematical {F}oundations of {Q}uantum {M}echanics}.
\newblock Princeton University Press, Princeton, NJ, 1955.

\bibitem{Kleene}
Stephen~Cole Kleene.
\newblock On notation for ordinal numbers.
\newblock \emph{J. Symbolic Logic}, 3:150--155, 1938.

\bibitem{Priest}
Graham Priest and Richard Sylvan.
\newblock Simplified semantics for basic relevant logics.
\newblock \emph{J. Philosophical Logic}, 21:217--232, 1992.

\bibitem{Lukasiewicz}
Jan {\L}ukasiewicz.
\newblock On {T}hree-{V}alued {L}ogic.
\newblock In L.~Borkowski, editor, \emph{Jan {\L}ukasiewicz. {S}elected
  {W}orks}, pages 87--88. North-Holland, Amsterdam, and PWN, Warsaw, 1970.

\bibitem{Gottwald}
Siegfried Gottwald.
\newblock \emph{A {T}reatise on {M}any {V}alued {L}ogics}.
\newblock Research Studies Press, 2001.

\bibitem{Reichenbach}
Hans Reichenbach.
\newblock \emph{Philosophic {F}oundations of {Q}uantum {M}echanics}.
\newblock Dover Publications, 2011.

\bibitem{Rey}
Georges Rey.
\newblock The {A}nalytic/{S}ynthetic {D}istinction.
\newblock In Edward~N. Zalta, editor, \emph{The Stanford Encyclopedia of
  Philosophy}. Metaphysics Research Lab, Stanford University,
  {https://plato.stanford.edu/archives/fall2018/entries/analytic-synthetic},
  2018.

\bibitem{Isham}
Chris~J. Isham.
\newblock Is {I}t {T}rue; or {I}s {I}t {F}alse; or {S}omewhere in {B}etween?
  {T}he {L}ogic of {Q}uantum {T}heory.
\newblock In William Demopoulos and Itamar Pitowsky, editors, \emph{Physical
  {T}heory and its {I}nterpretation}, pages 161--182. Springer, The
  Netherlands, 2006.

\bibitem{Humberstone}
Lloyd Humberstone.
\newblock \emph{The connectives}.
\newblock MIT Press, Cambridge, Mass., 2011.

\bibitem{Bergmann}
Merrie Bergmann.
\newblock \emph{An introduction to many-valued and fuzzy logic: semantics,
  algebras, and derivation systems}.
\newblock Cambridge University Press, 2008.

\bibitem{Piron}
Constantin Piron.
\newblock \emph{Foundations of {Q}uantum {P}hysics}.
\newblock W. A. Benjamin, Inc., 1976.

\bibitem{Halmos}
Paul~R. Halmos.
\newblock \emph{Introduction to {H}ilbert {S}pace and the {S}pectral {T}heory
  of {S}pectral {M}ultiplicity}.
\newblock Chelsea, New York, 1957.

\bibitem{Dickson}
Michael Dickson.
\newblock {Quantum Logic Is Alive $\wedge$ (It Is True $\vee$ It Is False)}.
\newblock \emph{Philosophy of Science}, 68(3):S274--S287, 2001.

\bibitem{Putnam68}
Hilary Putnam.
\newblock Is {L}ogic {E}mpirical{?}
\newblock In Robert~S. Cohen and Marx~W. Wartofsky, editors, \emph{Boston
  {S}tudies in the {P}hilosophy of {S}cience}, volume~5, pages 216--241. D.
  Reidel Publishing Company, 1968.

\bibitem{Bacciagaluppi}
Guido Bacciagaluppi.
\newblock Is {L}ogic {E}mpirical{?}
\newblock In Kurt Engesser, Dov~M. Gabbay, and Daniel Lehmann, editors,
  \emph{Handbook of {Q}uantum {L}ogic and {Q}uantum {S}tructures: {Q}uantum
  {L}ogic}, pages 49--78. Elsevier Science Publications, 2009.

\bibitem{Thomas}
George~B. Thomas and Ross~L. Finney.
\newblock \emph{Calculus and {A}nalytic {G}eometry, 9th edition}.
\newblock Dorling Kindersley Pvt Ltd, 2010.

\bibitem{Putnam75}
Hilary Putnam.
\newblock \emph{Mathematics, {M}atter and {M}ethod ({P}hilosophical {P}apers,
  {V}ol. 1)}.
\newblock Cambridge University Press, Cambridge, 1975.

\bibitem{Chiara}
Maria Luisa~Dalla Chiara, Roberto Giuntini, and Richard Greechie.
\newblock \emph{Reasoning in {Q}uantum {T}heory. {S}harp and {U}nsharp
  {Q}uantum {L}ogics}.
\newblock Springer Netherlands, 2004.

\bibitem{Griffiths}
Robert~B. Griffiths.
\newblock The {N}ew {Q}uantum {L}ogic.
\newblock \emph{Found Phys}, 44:610--640, 2014.

\bibitem{Dummett}
Michael Dummett.
\newblock \emph{Elements of {I}ntuitionism}.
\newblock Oxford University Press, Oxford, 1977.

\bibitem{Williamson}
Timothy Williamson.
\newblock \emph{Vagueness}.
\newblock Routledge, London, 1994.

\bibitem{Keefe}
Rosanna Keefe.
\newblock \emph{Theories of {V}agueness}.
\newblock Cambridge University Press, Cambridge, 2008.

\bibitem{Varzi}
Achille~C. Varzi.
\newblock Vagueness, logic and ontology.
\newblock \emph{The Dialogue. Yearbooks for Philosophical Hermeneutics},
  1:135--154, 2001.

\bibitem{Faye}
Jan Faye.
\newblock Copenhagen {I}nterpretation of {Q}uantum {M}echanics.
\newblock In Edward~N. Zalta, editor, \emph{The Stanford Encyclopedia of
  Philosophy}. {Metaphysics Research Lab, Stanford University,
  https://plato.stanford.edu/archives/win2019/entries/qm-copenhagen/}, 2019.

\bibitem{Trassinelli}
Martino Trassinelli.
\newblock Relational quantum mechanics and probability.
\newblock \emph{Found Phys}, 48:1092--1111, 2018.

\bibitem{Leggett}
Anthony~J. Leggett and Anupam Garg.
\newblock Quantum {M}echanics versus {M}acroscopic {R}ealism: {I}s the {F}lux
  {T}here when {N}obody {L}ooks?
\newblock \emph{Physical Review Letters}, 54(9):857--860, 1985.

\bibitem{Wigner}
Eugene~P. Wigner.
\newblock Remarks on the mind-body question.
\newblock In Irving~J. Good, editor, \emph{The Scientist Speculates}, pages
  284--302. William Heinemann, Ltd., London, 1961.

\bibitem{Frauchiger}
Daniela Frauchiger and Renato Renner.
\newblock Quantum theory cannot consistently describe the use of itself.
\newblock \emph{Nature Communications}, 9:3711, 2018.

\bibitem{Baumann}
Veronika Baumann and {$\check{\mathrm{C}}$}aslav Brukner.
\newblock Wigner’s friend as a rational agent.
\newblock {https://arXiv:1901.11274}, Jan 2019.

\bibitem{Stacey}
John~B. DeBrota and Blake~C. Stacey.
\newblock {FAQB}ism.
\newblock {arXiv:1810.13401 [quant-ph]}, 14 Apr 2019.

\bibitem{Fuchs}
Christopher~A. Fuchs, Nathaniel~David Mermin, and Ruediger Schack.
\newblock An {I}ntroduction to {Q}bism with an {A}pplication to the {L}ocality
  of {Q}uantum {M}echanics.
\newblock \emph{Am. J. Phys.}, 82:749--754, 2014.

\bibitem{Rovelli}
Carlo Rovelli.
\newblock Relational quantum mechanics.
\newblock \emph{International Journal of Theoretical Physics}, 35:1637--1678,
  1996.

\bibitem{Biagio}
Andrea~Di Biagio and Carlo Rovelli.
\newblock Stable {F}acts, {R}elative {F}acts.
\newblock \emph{Found. Phys.}, 51(30), 2021.

\bibitem{DeBrota}
John~B. DeBrota, Christopher~A. Fuchs, and Ruediger Schack.
\newblock {Respecting One's Fellow: QBism's Analysis of Wigner's Friend}.
\newblock \emph{Found. Phys.}, 50:1859--1874, 2020.

\bibitem{Bracewell}
Ronald Bracewell.
\newblock \emph{The {F}ourier {T}ransform \& {I}ts {A}pplications}.
\newblock McGraw-Hill Science/Engineering/Math, 1999.

\bibitem{Brukner}
{$\check{\mathrm{C}}$}aslav Brukner.
\newblock A {N}o-{G}o {T}heorem for {O}bserver-{I}ndependent {F}acts.
\newblock \emph{Entropy}, 20(350):1--10, 2018.

\bibitem{Bong}
Kok-Wei Bong, An{$\acute{\text{\i}}$}bal Utreras-Alarc{$\acute{\mathrm{o}}$}n,
  Farzad Ghafari, Yeong-Cherng Liang, Nora Tischler, Eric~G. Cavalcanti,
  Geoff~J. Pryde, and Howard~M. Wiseman.
\newblock A strong no-go theorem on the {W}igner’s friend paradox.
\newblock \emph{Nature Physics}, 16:1199--1205, 2020.

\bibitem{Proietti}
Massimiliano Proietti, Alexander Pickston, Francesco Graffitti, Peter Barrow,
  Dmytro Kundys, Cyril Branciard, Martin Ringbauer, and Alessandro Fedrizzi.
\newblock Experimental test of local observer independence.
\newblock \emph{Science Advances}, 5(eaaw9832), 2019.

\bibitem{Brown}
James~Robert Brown and Yiftach Fehige.
\newblock {Thought Experiments}.
\newblock In Edward~N. Zalta, editor, \emph{The Stanford Encyclopedia of
  Philosophy}. {Metaphysics Research Lab, Stanford University,
  https://plato.stanford.edu/archives/win2019/entries/thought-experiment/},
  2019.

\end{thebibliography}

\end{document}